\documentclass{article} 
\usepackage{iclr2024_conference,times}
\iclrfinalcopy

\usepackage{amsmath,amsfonts,bm}

\def\eqref#1{equation~\ref{#1}}

\def\1{\bm{1}}

\DeclareMathAlphabet{\mathsfit}{\encodingdefault}{\sfdefault}{m}{sl}
\SetMathAlphabet{\mathsfit}{bold}{\encodingdefault}{\sfdefault}{bx}{n}

\newcommand{\R}{\mathbb{R}}

\usepackage{hyperref}
\usepackage{url}
\usepackage{graphicx}
\usepackage{array}
\usepackage{comment}
\usepackage{algorithm}
\usepackage{algpseudocode}
\usepackage{multirow}
\usepackage{listings}

\usepackage{xcolor}

\definecolor{codegreen}{rgb}{0,0.6,0}
\definecolor{codegray}{rgb}{0.5,0.5,0.5}
\definecolor{codepurple}{rgb}{0.58,0,0.82}
\definecolor{backcolour}{rgb}{1.0,1.0,1.0}

\lstdefinestyle{mystyle}{
    backgroundcolor=\color{backcolour},   
    commentstyle=\color{codegreen},
    keywordstyle=\color{magenta},
    numberstyle=\tiny\color{codegray},
    stringstyle=\color{codepurple},
    basicstyle=\ttfamily\footnotesize,
    breakatwhitespace=false,         
    breaklines=true,                 
    captionpos=b,                    
    keepspaces=true,                 
    numbers=left,                    
    numbersep=5pt,                  
    showspaces=false,                
    showstringspaces=false,
    showtabs=false,                  
    tabsize=2
}

\title{Force Field Optimization by End-to-end Differentiable Atomistic Simulation}

\author{Abhijeet S. Gangan \\
        Department of Civil and Environmental Engineering \\
        University of California, Los Angeles \\
        CA 90095, USA \\
	\texttt{abhijeetgangan@g.ucla.edu} \\
 	\And
        {Ekin Dogus Cubuk} \\
        Google DeepMind \\
        Mountain View \\ 
        CA, USA \\
	\texttt{cubuk@google.com} \\
  	\And
	{Samuel S. Schoenholz} \\
        OpenAI, San Francisco\\ 
        CA, USA \\
 	\And
	{Mathieu Bauchy} \\
        Department of Civil and Environmental Engineering \\
        University of California, Los Angeles \\
        CA 90095, USA \\
	\texttt{bauchy@ucla.edu} \\
  	\And
	{N. M. Anoop Krishnan} \\
        Department of Civil Engineering,\\
        Yardi School of Artificial Intelligence,\\
        Indian Institute of Technology Delhi, \\
        New Delhi, India 110016 \\
	\texttt{krishnan@iitd.ac.in} \\}

\begin{document}
\maketitle

\begin{abstract} 
The accuracy of atomistic simulations depends on the precision of force fields. Traditional numerical methods often struggle to optimize the empirical force field parameters for reproducing target properties. Recent approaches rely on training these force fields based on forces and energies from first-principle simulations. However, it is unclear whether these approaches will enable capturing complex material responses such as vibrational, or elastic properties. To this extent, we introduce a framework, employing inner loop simulations and outer loop optimization, that exploits automatic differentiation for both property prediction and force-field optimization by computing gradients of the simulation analytically. We demonstrate the approach by optimizing classical Stillinger-Weber and EDIP potentials for silicon systems to reproduce the elastic constants, vibrational density of states, and phonon dispersion. We also demonstrate how a machine-learned potential can be fine-tuned using automatic differentiation to reproduce any target property such as radial distribution functions. Interestingly, the resulting force field exhibits improved accuracy and generalizability to unseen temperatures than those fine-tuned on energies and forces. Finally, we demonstrate the extension of the approach to optimize the force fields towards multiple target properties. Altogether, differentiable simulations, through the analytical computation of their gradients, offer a powerful tool for both theoretical exploration and practical applications toward understanding physical systems and materials. 
\end{abstract}

\section{Introduction}
Atomistic simulations offer direct insight into atomic movements, which, in turn, dictate the macroscopic properties of materials~\citep{frenkel2023understanding}. The force field, responsible for predicting the total energy of a system based on atomic coordinates, serves as the cornerstone of these simulations, determining their accuracy and reliability~\citep{torrens2012interatomic}. The primary objective of a force field is to replicate atomic motions observed in first-principle simulations as closely as possible while faithfully reproducing experimentally observed properties as well. Traditionally, force field development has focused on fitting parameters to match experimental properties, such as elastic moduli, lattice parameters of crystalline systems, density, and other complex characteristics~\citep{torrens2012interatomic,pedone2022interatomic,purja2009development,wang2018new, EXP_SIM, Learning_Force_Field, HybridClassicalML}. However, with the emergence of machine-learned force fields (MLFFs), recent efforts have shifted toward training models directly on energy and force data (or other higher order properties such as stresses) derived from first-principle computations~\citep{noe2020machine,deringer2019machine} or learning them from the observables~\citep{thaler2021learning,bishnoi2023discovering,stochastic_particle_trajectories,IPSuite}. Both approaches share the ultimate goal of optimizing force field parameters by minimizing the discrepancy between simulations and ``ground truth'' force and energy values or other properties such as trajectories. Thus, the success of these methodologies critically hinges on the precise computation of the gradients with respect to the simulated quantities.

Efficient and accurate calculation of derivatives is crucial in atomistic simulations, whether for force field optimization or for computing gradient-based properties integral to these simulations~\citep{jaxmd2020,brenner2023scientific}. In force field optimization, the derivatives of properties with model parameters are essential. However, this task becomes particularly challenging as the number of model parameters increases. One common approach to approximate derivatives is the finite difference method. While straightforward, this method becomes computationally prohibitive in high-dimensional spaces, as the number of function evaluations scales with the dimension. Additionally, the finite difference method is susceptible to truncation and round-off errors, making it impractical for large-scale applications.

Alternatively, symbolic differentiation can be employed~\citep{durrbaum2002comparison}, where derivatives are computed analytically using computer algebra systems such as Mathematica and SymPy~\citep{meurer2017sympy}. Although this approach yields precise analytical derivatives, the resulting expressions can grow exponentially complex due to the tree-like structure of operations and the repeated application of the chain rule. Manual differentiation, where gradients are computed by hand and then coded, is another option; however, this method is often time-consuming and prone to human error~\citep{strikwerda2004finite}. Despite these challenges, traditional atomistic simulation packages, such as LAMMPS~\citep{LAMMPS}, often rely on manually coded derivatives (forces), which underscores the importance of reliable and efficient derivative computation in these simulations.    

The machine learning community has long utilized efficient algorithms such as back-propagation to compute gradients, a technique that is a specialized form of a broader method known as Automated or Algorithmic Differentiation (AD). AD enables the efficient computation of analytical gradients, which is critical for optimizing complex models. Libraries such as JAX and PyTorch have harnessed the power of AD to create end-to-end differentiable simulation codes, which are increasingly being applied to the simulation of physical systems~\citep{fliudAD,thangamuthu2022unravelling,bishnoibrognet, Learning_pair_potential} and a wide range of optimization and control problems \citep{CHEMAD, AEROAD,hu2019difftaichi,chen2022daxbench,AthermalDesign}. On an atomistic simulations front, JAX-MD is an actively maintained package that leverages these capabilities. There have also been other attempts to optimize geometries~\citep{axelrod2022learning}, control~\citep{wang2020differentiable}, sampling~\citep{schwalbe2021differentiable}, and learning pair-potentials~\citep{wang2023learning}. Despite the wider adoption of AD for several optimization problems, there have been limited efforts to exploit its use for direct optimization of force fields for any targeted property of interest~\citep{brenner2023scientific}. 

Here, we apply AD as implemented in JAX-MD to optimize force fields for various phases of silicon, aiming to closely match the static and dynamic properties predicted by first-principles simulations such as elastic tensors, vibrational density of states, and radial distribution functions (RDF). We show that AD can be used to identify the optimal parameters for both classical and MLFFs in just 4-5 iterations. Further, we also show that the approach could be easily extended to multi-objective optimizations to tune force fields towards replicating multiple target properties.
Our results suggest AD has the potential to revolutionize atomistic simulations by enabling the computation of the gradients of simulation---a feature that can have implications beyond force field optimization.



\section{Methodology}
\subsection{Dataset generation}
\label{datagen}
To demonstrate the capabilities of differentiable atomistic simulation for force-field optimization, we rely on first-principles calculations to generate the ground truth. Without the loss of generality, we focus on silicon (Si), a material that has been extensively studied in the past few decades~\cite{sastry2003liquid}, exhibiting both amorphous and crystalline forms. Several structural, physical, and mechanical properties, such as elastic tensor, phonon density of states, and radial distribution functions, of the material, are computed using first-principles calculation---the framework of plane-wave density functional theory (DFT) as implemented in the Vienna Ab Initio Simulation Package (VASP)~\citep{VASP1, VASP2} is used. The standard VASP pseudo-potential (PSP) based on the projector augmented wave (PAW)~\citep{PAW1, PAW2} is used for Si. For the exchange-correlation functional, the Perdew-Burke-Ernzerhof (PBE) with generalized gradient approximation (GGA)~\citep{PBE} is employed. The Si PSP has the 4 valence electrons with the $3s^2 3p^2$ as the configuration. All the simulations are carried out using periodic boundary conditions (PBC).

The phonon and elastic constants are computed at 0 K through molecular statics employing energy minimization. For computing phonon and elastic constants, a plane-wave cutoff of $520$ eV is used with tighter energy and force convergence criterion of $10^{-9}$ eV and $10^{-7}$ eV/{\AA}, respectively. This along with a $\Gamma$-centered \textit{k}-point mesh of $12 \times 12 \times 12$ ensures accurate convergence of these properties. Gaussian smearing of $0.05$ eV is employed and spin-polarization is not considered as the Si phases simulated have a non-magnetic ground state. Elastic constants are computed on a unit cell with 8 Si atoms. For phonon calculations, a super-cell that is $2\times2\times2$ the unit-cell yields converged phonon frequencies. The \textit{k}-point mesh is adjusted for the super-cell to keep the same \textit{k}-point density. Generation of displacement and post-processing of the force constant data is performed with \texttt{phonopy} ~\citep{phonopy}.

For evaluating the performance of the potential in molecular dynamics simulations, finite temperature \textit{ab initio} molecular dynamics (AIMD) is employed to generate the ground truth. For AIMD simulations, a lower cutoff of 400 eV is used along with an energy convergence criterion of $10^{-6}$ eV. A super-cell that is $2\times2\times2$ the unit-cell is used as a starting point. $2\times2\times2$ $\Gamma$-centered \textit{k}-point mesh is used to obtain accurate forces as Silicon in the liquid phase exhibits a metallic nature, that is, the density of states at Fermi energy is non-zero. A time step of $1$ fs is used for integration of the equation of motion. The temperature is controlled using the Nosé–Hoover thermostat \citep{NOSE2,NOSE1}. The crystal is equilibrated at 2000 K and the initial 5000 steps (5 ps) for melting are discarded to avoid any remnants of the crystal structure. The radial distribution function (RDF) obtained from the simulation is used as the ground truth representing the structure of the Si system. The averaging is done based on the steps used in the differentiable simulation setup. The process is repeated for additional temperatures of 1000 K and 3000 K and a larger cell (that is, $3\times3\times3$ the unit-cell). These structures are maintained as the test set to evaluate the generalizability of the force field to unseen temperatures. For the training MLFFs, we employ an existing dataset \citep{SiGNNdataset}
that provides the energy and forces of Si for different configurations generated using DFT. \textcolor{black}{We use the energy and force data at 300K, 600K, and 900K of the cubic phase and that of the liquid phase Silicon. The training, validation, and testing set consists of 2578, 859, and 860 energy and force data points respectively.} 

\subsection{Differentiable simulation}
Atomistic simulations are carried out using the end-to-end differentiable package, namely, JAX-MD. JAX-MD allows computing analytical gradients to model parameters that are not available in traditional MD packages like LAMMPS. To the best of our knowledge, JAX-MD is the only package that allows differentiable simulation of atomistic systems. We do, however, use LAMMPS to check the consistency of the classical simulations. The parameters of the simulation like the timestep, temperature, etc are kept similar to those in the \textit{ab initio} simulation mentioned above to avoid inconsistencies during the simulation. Additional details on how several properties such as elastic moduli are computed an example code snippet are mentioned in the supplementary materials.

\subsection{Training details}
For the \texttt{scipy} optimization routines the max iteration is set to 200 and the convergence tolerance is made tighter. Other settings are kept as default. A cutoff of $3$ \AA \ was used to generate graphs from the atomic coordinates. The GNN is repeated for $k=2$ messages and the MLP for the update, message functions, and the decoder $64$ neurons with $2$ layers is used. For training the GNN we used the Adam optimizer with a learning rate of $1e-4$. The fine-tuning process was carried out for a total of 2000 epochs with a batch size of 64. For the evaluation, we save the model with the lowest validation error. 

\label{Hyperparameters}
%
Loss functions govern the nature of the model training and the final optimized solutions. For the tensor property of rank two, such as the elastic constants $C_{ij}$ and force constants $H_{ij}$, Frobenius norm is used as the loss function given by:
\begin{equation}
    \label{eq:fro1}
    L_{F} = \sqrt{\sum_{i}\sum_{j}|Y_{ij} - \hat{Y}_{ij}|^{2}}.
\end{equation}

where, $Y_{ij}$ is true value of the property, with the hat ( $\hat{}$ ) representing the predicted counterpart. For the multi-objective optimization, we combine the Frobenius norm for two properties and use the weights $w_1$ and $w_2$ which determine the scaling of the loss values and are hyperparameters.
\begin{equation}
\label{eq:fro2}
    L_{T} = w_{1} \sqrt{\sum_{i}\sum_{j}|Y_{ij} - \hat{Y}_{ij}|^{2}} +  w_{2}\sqrt{\sum_{i}\sum_{j}|Z_{ij} - \hat{Z}_{ij}|^{2}}
\end{equation}
For fine-tuning the message passing graph neural network (MPGNN), we use the mean-squared error \eqref{eq:RMSE} between energies $E$ and forces $\vec{F}$ for different configurations $i \in \textsc{I}$. 

\begin{equation}
\label{eq:woRDF}
  L = \frac{1}{n} \sum_{i=1}^{n} (E_{i} - \hat{E}_{i})^2 + \frac{1}{n} \sum_{i=1}^{n} (\vec{F}_{i} - \hat{\vec{F}}_{i})^2
\end{equation}

For finetuning the MPGNN on radial distribution functions, we add a term $\chi^{2}$ in the loss function to measure the difference between the $g(r)$. We add a hyperparameter $w$ for scaling the $\chi^{2}$, which is set to a value of $2000$ in the present work.
\begin{equation}
\label{eq:RDF}
  L = \frac{1}{n} \sum_{i=1}^{n} (E_{i} - \hat{E}_{i})^2 + \frac{1}{n} \sum_{i=1}^{n} (\vec{F}_{i} - \hat{\vec{F}}_{i})^2 + w{\chi}^{2}
\end{equation}
where $\chi^{2}$ is defined as:
\begin{equation}
   \chi^2 = \frac{\sum_{i}^{n} [g(r) - \hat{g}(r)]^2}{\sum_{i}^{n} [\hat{g}(r)]^2}
\end{equation}

The performance of the model is evaluated on the root mean squared error (RMSE), $L^1$ (mean absolute error, MAE), and $L^2$ (mean squared error, MSE) norms of the predicted quantity such as energy, force or radial distribution function as follows. Note that this metric is primarily used for the MPGNN models. 
\begin{equation}
\label{eq:RMSE}
RMSE = \sqrt{\frac{1}{n} \sum_{i=1}^{n} (y_i - \hat{y}_i)^2}; \quad L^1 = \frac{1}{n} \sum_{i=1}^{n} |y_i - \hat{y}_i|; \quad L^2 = \frac{1}{n} \sum_{i=1}^{n} (y_i - \hat{y}_i)^2
\end{equation}



\subsection{Interaction potentials for Silicon}
\subsubsection*{\textit{Stillinger-Weber potential}}
\label{SWP}
We use the Stillinger-Weber (SW) potential to describe the atomistic behavior of bulk Silicon (Si) \cite{SWREF}. The potential energy of all the atoms in a bulk Si system is given by:

\begin{equation}
U_{Total} = \sum_{i}\sum_{j>i}\phi_2(r_{ij}) + \sum_{i}\sum_{j\neq{i}}\sum_{k>j} \phi_3(\vec{r}_{ij},\vec{r}_{ik},\theta_{ijk})
\end{equation}

where the $\phi_2$ and $\phi_3$ are the two and three-body interaction terms that have the following functional form:
\begin{equation}
\label{eq:sw2}
\phi_2(r_{ij}) = A_{ij}\epsilon_{ij}\left [B_{ij}\left(\frac{\sigma_{ij}}{r_{ij}}\right)^{p_{ij}} - \left(\frac{\sigma_{ij}}{r_{ij}}\right)^{q_{ij}}\right]\exp{\left(\frac{\sigma_{ij}}{r_{ij} - a_{ij}\sigma_{ij}}\right)}
\end{equation}
the first term in \ref{eq:sw2} repulsive part while the second term is the attractive part. 
\begin{equation}
\phi_3(\vec{r}_{ij},\vec{r}_{ik},\theta_{ijk}) = \lambda_{ijk}\epsilon_{ijk} [\cos{\theta_{ijk}} - \cos{\theta_{0ijk}}]^{2}\exp{\left(\frac{\gamma_{ij}\sigma_{ij}}{r_{ij} - a_{ij}\sigma_{ij}}\right)}\exp{\left(\frac{\gamma_{ik}\sigma_{ik}}{r_{ik} - a_{ik}\sigma_{ik}}\right)}
\end{equation}

The parameters $\epsilon$ and $\sigma$ determine the energy and distance scaling respectively. The $A$ and $B$ parameter is chosen so that the minimum of the $\phi_2(r_{ij})$ i.e $r_{0}=2^{1/6}\sigma$ has the energy $-\epsilon$. Here $r_0$ is the equilibrium bond length between two atoms and $r_{c}=a\sigma$ is the cutoff for the potential beyond which the potential energy is zero. $\theta_{0ijk}$ is the equilibrium bond angle between triplets of atoms. The $\lambda$ controls the strength of the three-body term and .$\gamma$ controls the rate at which the three-body term goes to zero. The $p$ and $q$ are free parameters that control the curvature of the repulsive part and the attractive part respectively. The $\theta_{0ijk} = 109.47^{\circ}$ is the tetrahedral bond angle which ensures that the diamond structure is in the minimum energy (ground) state.

\begin{table}
\caption{Original parameters of the SW potential for Si.}
    \centering
    \begin{tabular}{|c|c|c|c|}
    \hline
    Params& Values& Params&Values\\
    \hline
    $\epsilon$ (eV)&  2.16826&  $A$& 7.049556277\\
     $\sigma$  (\AA)&  2.0951&  $B$& 0.6022245584\\
    $a$&  1.80&  $p$& 4.0\\
    $\lambda$&  21.0&  $q$&  0.0\\
    $\gamma$&  1.20&  & \\
    \hline
    \end{tabular}
    \label{tab:SWparameters}
\end{table}

\subsubsection*{\textit{Environment dependent interatomic potential}}
\label{EDIP}
Environment-dependent interatomic potential (EDIP) is a bond order potential originally used for bulk silicon \cite{EDIP1,EDIP2}. The pairwise potential energy $U_i$ depends on the local environment of the atom. This is taken into account by bond order functions which depend on the local coordination number. This way the potential is able to capture the different types of chemical bonding for different coordinations for different phases of Silicon. 

The potential energy is the sum of two-body and three-body terms and is given by -

\begin{equation}
  U_i = \sum_{j\neq i} \phi_{2}(r_{ij}) + \sum_{j\neq i}\sum_{k\neq i,k>j}\phi_{3}(\vec{r}_{ij},\vec{r}_{ik},\theta_{ijk})
\end{equation}

where $Z_{i}$ the effective co-ordination number is defined by -

\begin{equation}
  Z_{i} = \sum_{m\neq i} f(r_{im})
\end{equation}

The sum is carried over the neighbors of the atom. Here $f(R_{im})$ is the cutoff function which drops smoothly to $0$ from $1$ from $a$ to $c$. Here $a$ and $c$ are outer and inner cutoffs. This function depends only on the separation between the atoms $r$.

\begin{equation}
f(r) = \begin{cases} 
1 & r \leq c \\
\exp(\frac{\alpha}{1-x^{-3}}) & c < r < a \\
0 & a \geq r 
 \end{cases}
\end{equation}

where the $x$ takes the form -

\begin{equation}
x = \frac{r-c}{a-c}
\end{equation}

The two-body potential is given by -
\begin{equation}
  \phi_{2}(r,Z) = A \left[(\frac{B}{r})^{\rho} - p(Z)\right]\exp{\left(\frac{\sigma}{r-a}\right)}
\end{equation}

Here $p(Z)$ is the Gaussian function which depends on the effective local coordination $Z$.

\begin{equation}
p(Z) = \exp({-\beta Z^2})
\end{equation}

The three-body term is given by -
\begin{equation}
\phi_{3}(\vec{r}_{ij},\vec{r}_{ik},Z_{i}) = g(r_{ij})g(r_{ik})h(l_{ijk},Z_{i})
\end{equation}

where $l_{ijk} = \cos(\theta_{ijk}) = \vec{r}_{ij} \cdot \vec{r}_{ik}/r_{ij}r_{ik}$

The radial part $g(r)$ is given by -
\begin{equation}
  g(r) = \exp{\left(\frac{\gamma}{r-a }\right)}
\end{equation}

The angular part of the function $h(l,Z)$ depends on the effective local coordination and is described as -

\begin{equation}
  h(l,Z) = \lambda[(1 - \exp(-Q(Z)(l + \tau(Z))^2)) + \eta Q(Z)(l + \tau(Z))^2]
\end{equation}

where $Q(Z) = Q_{0}\exp(-\mu Z)$ and the function $\tau(Z)$ is given by -

\begin{equation}
  \tau(Z) = u_1 + u_2(u_3 \exp(-u_{4}Z) - \exp(-2u_{4}Z))
\end{equation}

$A$ and $\lambda$ determine the energy scales of the two-body and three-body terms respectively. $\sigma$ determines the distance scale between neighbors while $B$ and $\gamma$ are parameters of the repulsive two-body term and radial three-body term respectively. $\rho$ is the exponent for the repulsive part of the two-body term while $\beta$ is a parameter for the two-body bond order function $p(Z)$. The two functions also depend on $Z$. $u_1, u_2, u_3$ and $u_4$ are parameter for the three-body bond order function $\tau(Z)$. $Q_0$ and $\mu$ are parameters for the three-body bond order function Q(Z) while $\eta$ is a parameter for $h(l,Z)$. $a$ and $c$ are the outer and inner cutoff for the function $f(r)$ and $\alpha$ the exponential scaling parameter.

\begin{table}
    \caption{Original parameters of the EDIP potential for Si.}
    
    \centering
    \begin{tabular}{ |c|c|c|c| }
    \hline
    Params& Values& Params&Values\\
    \hline
    $u_{1}$&  -0.165799&  $\alpha$& 3.1083847\\
    $u_{2}$&  32.557&  $A$& 7.9821730\\
    $u_{3}$&  0.286198&  $\lambda$& 1.4533108\\
    $u_{4}$&  0.66&  $B$& 1.5075463\\
    $\rho$&  1.2085196&  $\gamma$& 1.1247945\\
    $\eta$& 0.2523244& $\sigma$&0.5774108\\
    $Q_0$& 312.1341346& $c$&2.5609104\\
    $\mu$& 0.6966326& $a$&3.1213820\\
    $\beta$& 0.0070975& &\\
    \hline
    \end{tabular}
    \label{tab:EDIPparameters}
\end{table}
\subsubsection*{\textit{Machine learned interatomic potential}}
\label{GNNmodel}
Consider a system with $N$ particles. The system is represented as a graph $G = (V, E)$ where:
\begin{itemize}
    \item $V = \{v_1, v_2, \ldots, v_N\}$ is the set of nodes representing particles.
    \item $E \subseteq V \times V$ is the set of edges representing particle interactions.
\end{itemize}

Each node $v_i$ has a feature vector $ \mathbf{h}_i$, and each edge $(v_i, v_j)$ has a feature vector $\mathbf{e}_{ij}$.

The GNN consists of multiple layers, each performing message passing and updating node features.

\begin{equation}
\mathbf{m}_i^{(k)} = \sum_{j \in \mathcal{N}(i)} F_{\text{msg}}^{(k)}(\mathbf{h}_i^{(k)}, \mathbf{h}_j^{(k)}, \mathbf{e}_{ij})
\end{equation}

Here, $\mathbf{m}_i^{(k)}$ is the message for node $i$ at layer $k$, $\mathbf{h}_i^{(k)}$ and $\mathbf{h}_j^{(k)}$ are the node features, and $\mathbf{e}_{ij}$ is the edge feature.

\begin{equation}
\mathbf{h}_i^{(k+1)} = F_{\text{update}}^{(k)}(\mathbf{h}_i^{(k)}, \mathbf{m}_i^{(k)})
\end{equation}

Here, $\mathbf{h}_i^{(k+1)}$ is the updated feature for node $i$ at layer $k+1$.

The global state is updated based on the final node features:

\begin{equation}
\mathbf{g} = \text{Aggregate}(\{\mathbf{h}_i^{(K)}\})
\end{equation}

Here, $\mathbf{g}$ is the global state obtained by aggregating the final node features $\mathbf{h}_i^{(K)}$.

The energy $U$ is computed using a decoder applied to the global state:

\begin{equation}
U = D(\mathbf{g})
\end{equation}

Here, $D$ is the function that maps the global state to the system's potential energy and $K$ is the total number of message-passing steps. The functions $F_{update}$, $F_{msg}$, and $D$ are neural networks and their weights are learned during the training.

\subsection{Molecular simulation setup}
\label{MDsetup}
The elastic constants, $C_{ijkl}$ (see Eq.\ref{eq:ec_hc}), are computed as the double derivative of the energy with strains using the \texttt{elasticity.py} module in JAX-MD. For computing the force constants, the Hessian, $H_{ij}$ (see Eq.\ref{eq:ec_hc}), is computed using the auto-grad in JAX and then post-processed by  the\texttt{phonopy} package. The details of the computation are described in the Supplementary along with the computation of thermodynamics properties.
\begin{equation}
\label{eq:ec_hc}
    C_{ijkl} = \frac{1}{V}\frac{\partial^{2} U}{\partial \epsilon_{ij} \partial \epsilon_{kl}}; \quad H_{ij} = \frac{\partial^2 U}{\partial R_{i} \partial R_{j}}
\end{equation}


For simplicity, we use the Voigt notation for representing the elastic tensors in the rest of the manuscript with reduced subscripts as $C_{ij}$. For the molecular dynamics simulation for computing the RDF and training the GNN model, we use both the NVE and NVT ensembles with a time step of 1 fs and with a total of 200 steps with a temperature of 2000 K. The gradient is then computed by differentiating through this simulation. In the testing part, this simulation is repeated for 5000 steps. The results presented are for the NVE ensemble as they are almost identical to the NVT results but are faster in terms of simulation time. All the simulations use periodic boundary conditions.

\section{Results}
\subsection{Differentiable simulation}
\begin{figure}[!htbp]
    \centering
    {\includegraphics[width=1.0\textwidth]{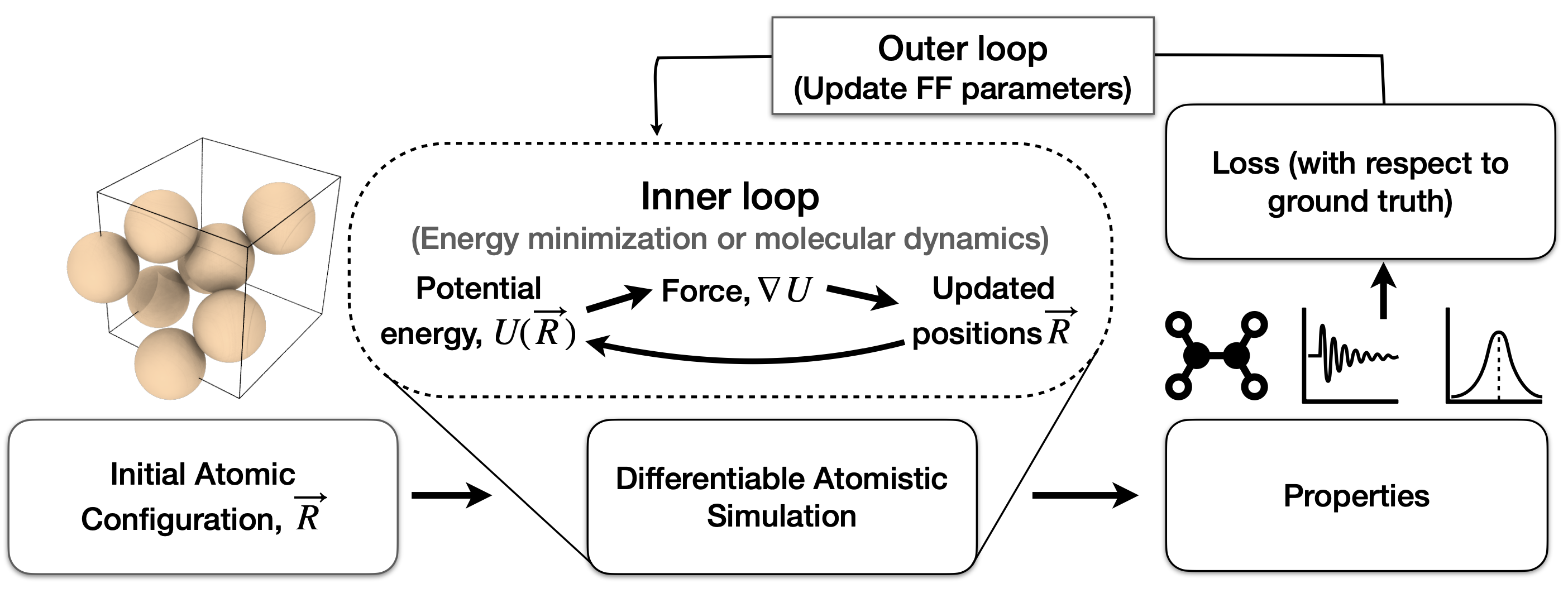}}
    \vspace{-0.25in}
\caption{Schematic illustrating differentiable simulation for the optimization problem.} 
\label{fig:fig1}
\vspace{-0.1in}
\end{figure}

First, we briefly discuss how the differentiable atomistic simulation framework enables the optimization of the force-field parameters. Fig.~\ref{fig:fig1} shows the overall pipeline employed in the present work. The objective of the optimization problem is to obtain the optimal force-field parameters $\mathbf{\Theta}^{*}$ associated with the potential energy function $\mathbf{U}(\mathbf{{R}}, \mathbf{\Theta})$ by minimizing the function $f$, defined as:
\begin{equation}
    \mathbf{\Theta}^{*} = \text{argmin}\ f(\mathbf{R}, \mathbf{\Theta})
\end{equation}

where, $\mathbf{R} \in \R^{N\times d}$ represents the position vectors associated with $N$ atoms in $d$ dimensions, $\Theta = \{\theta_{1}, \theta_{2}, ..., \theta_{k}\}, \theta_{k} \in \R$ are the parameters of the force-field, and $f$ is the loss function representing the difference between the ground truth and the predictions from the simulation. Note that $\mathbf{U}$ represents the total energy of the system and the atom-level forces are obtained as the gradient of $\mathbf{U}$. It is worth noting that the loss function is on the properties that can be obtained through simulations only. For instance, despite having accurate interatomic potentials, material properties such as elastic modulus, and phonon density of states can be computed only by performing an atomistic simulation. Here, we optimize the parameters of the force field for a given target property using a differentiable simulation. 

Specifically, the framework consists of an inner simulation loop and an outer optimization loop (see Fig.~\ref{fig:fig1}). The inner simulation loop carries out traditional atomistic simulation (energy minimization or molecular dynamics) to compute the property. Then, the error between the property computed through the inner simulation loop and the ground truth is computed. Following this, an outer back-propagation loop updates the parameters of the force field to minimize the error. Note that the parameter update is performed by computing the direction in which loss decreases through gradient descent. Thus, differentiable simulation allows the computation of the gradients of a simulation directly with the potential parameters or other input parameters. This approach enables one to bypass the expensive numerical differentiation approaches such as finite difference which demands a large number of data points while still having numerical approximation errors. 

\subsection{Elastic constants}
\label{re_elast}

\begin{table}
\caption{Original and optimized parameters for SW and EDIP force fields by minimizing the error on elastic constant using different optimization algorithms.}
  \centering
  \begin{tabular}{|c|c|c|c|c|} \hline 
    \multirow{2}{*}{Force field params} &\multicolumn{2}{c|}{SW} & \multicolumn{2}{c|}{EDIP}\\ \cline{2-5}
    
     & $\epsilon$ (eV) &  $\lambda$ & A (eV) &  $\lambda$ (eV) \\
    \hline
    Original & 2.16826 & 21.0 & 7.9821730 & 1.4533108 \\
    BFGS& 1.935160956627 &  33.786738332455 & 7.191596385156 &  1.457774753403 \\
    CG& 1.935160954811 &  33.786738320322& 7.191596392526 &  1.457774758363  \\
    NM& 1.935159113520 &  33.786797615075 & 7.191619415626 &  1.457752548290\\
    \hline 
  \end{tabular}
  \label{tab:Elastparam}
  \vspace{-0.25in}
\end{table}

\begin{table}
 \caption{Comparison of elastic constants $C_{ij}$ for original and optimized parameters of SW and EDIP potential along with DFT values.}
  \centering
  \begin{tabular}{|c|c|c|c|c|} \hline 
      \multicolumn{2}{|c|}{Force field} & $C_{11}$ (GPa) &  $C_{12}$ (GPa) & $C_{44}$ (GPa)\\
    \hline
    \multirow{2}{*}{SW} & Original & 151.4131 & 76.4154 &  112.8927 \\
    & Optimized & 162.3052 & 54.6155 & 140.8832 \\
    \hline
    \multirow{2}{*}{EDIP} & Original & 172.0389 & 64.6745 &  145.5683 \\
    &Optimized & 162.3106 & 54.6135 & 140.8778 \\
    \hline
    \multicolumn{2}{|c|}{DFT} & 153.2141 & 56.7969 & 149.6653 \\ \hline
  \end{tabular}
  \label{tab:Elast}
  \vspace{-0.25in}
\end{table}

Now, we demonstrate how differentiable simulation can be employed to optimize the force-field parameters, both classical and machine-learned, towards a target property. We start with the optimization of the classical SW and EDIP potential parameters for Si to match the elastic constants from DFT. Details of the potentials and elastic constants' computation from DFT are provided in the Methodology section. Since the elastic tensor has multiple components, we use the Frobenius norm as the loss function to minimize \eqref{eq:fro1} where $Y_{ij}$ and $\hat{Y}_{ij}$ are matrix elements of $C_{ij}$ computed from forcefields in JAX-MD and DFT, respectively. Note that in this case, the elastic tensor Eq. \ref{eq:ec_hc} computation in the inner loop and the backpropagation to minimize the Frobenius norm in the outer loop is performed using automatic differentiation. Thus, in addition to the optimization, the elastic tensor is computed by computing the Hessian of the energy analytically with strains using automatic differentiation. This approach reduces the computational cost associated with multiple simulation steps and avoids any numerical errors associated with the finite difference method.

To optimize the force-field parameters, we consider three different optimization algorithms, that is, Conjugate gradient (CG), Broyden–Fletcher–Goldfarb–Shanno (BFGS), and Nelder–Mead (NM). CG and BFGS are gradient-based optimization algorithms that can utilize gradients computed from JAX-MD directly, whereas NM is a gradient-free minimizer. For each minimizer, optimization is performed with four different initial conditions to obtain the statistics. Although there are several parameters for these potentials, without the loss of generality, we choose $\epsilon$ and $\lambda$ for SW and $A$ and $\lambda$ for EDIP as they govern the scaling of the two and three-body terms, respectively \cite{EDIP2,SWREF}. The other parameters are kept fixed and the default values of the potentials are given in table \ref{tab:SWparameters}. The approach could be used to optimize other parameters also by simply computing their gradients as well. 


\begin{figure}[!htbp]
    \centering
    {\includegraphics[width=0.75\textwidth]{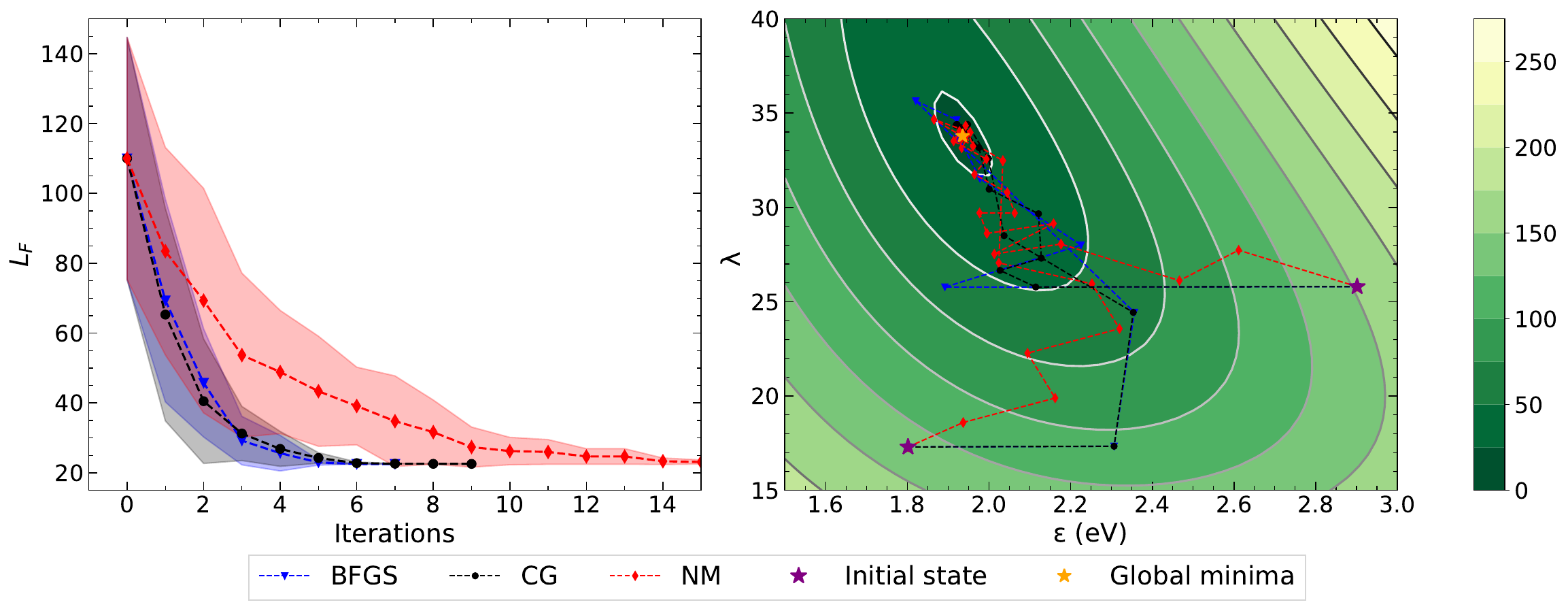}}
    {\includegraphics[width=0.75\textwidth]{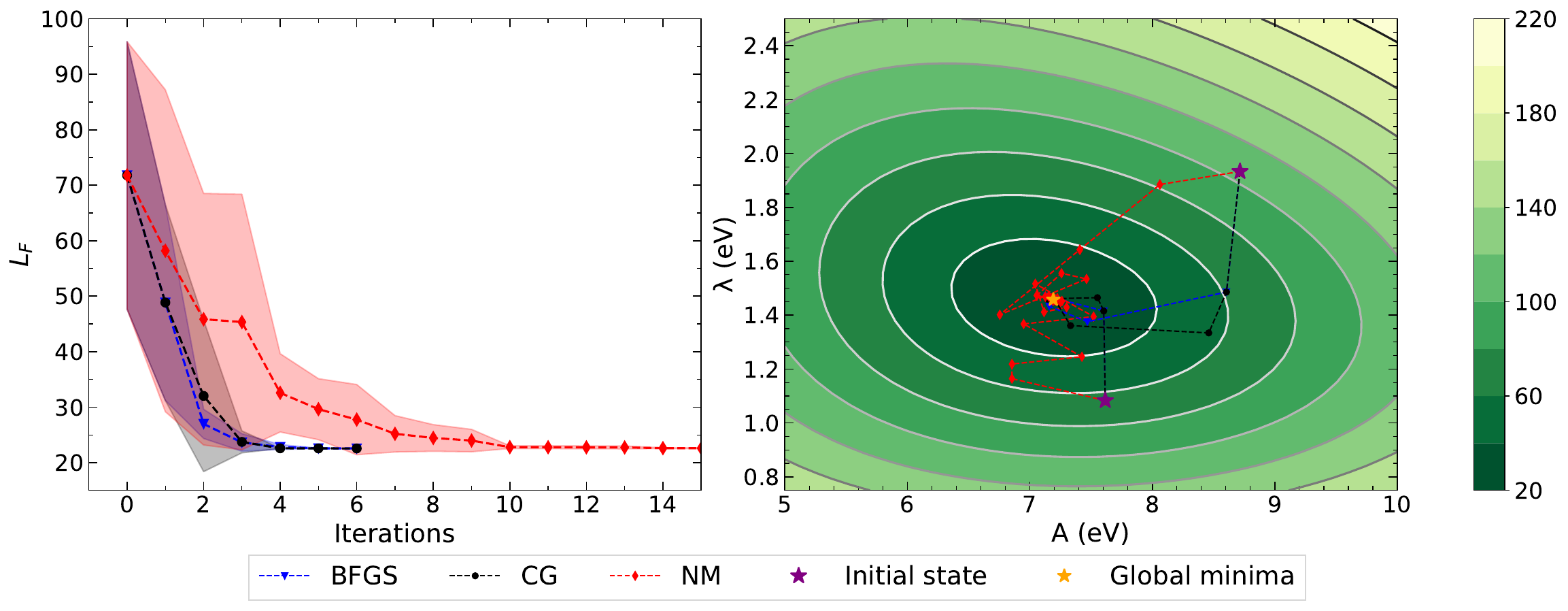}}
    \caption{The learning curve and the optimization pathway in the parameter space with different optimizers for SW (top) and EDIP (bottom) potentials. The contour values are the values of the loss function at each parameter. } 
    \label{fig:fig2}
    \vspace{-0.2in}
\end{figure}

Figures~\ref{fig:fig2}(a) and (b) show the loss curve for the SW and EDIP potentials. We observe that the gradient-based algorithms (BFGS and CG) saturate faster than NM for both SW and EDIP. Further, both BFGS and CG exhibit similar behavior for both potentials with loss saturating at 6 and 4 iterations for SW and EDIP potential, respectively. The optimized parameters for SW and EDIP are shown in Table \ref{tab:Elastparam}. We observe that the optimized values of parameters obtained by all the minimizers are comparable while being significantly different from the original values. To evaluate this behavior further, we plot the loss landscape for the potential parameters and project the minimization trajectory onto the landscape. Figures~\ref{fig:fig2}(c) and (d) show the loss landscape for SW and EDIP potentials, respectively, for two random initial states. We observe that the loss landscape exhibits a convex nature. Both BFGS and CG reach the global minima in ~4 steps irrespective of the initial condition, consistent with the loss curve. Thus, the differentiable can find the global minimum for the optimization in an accelerated fashion.

We also evaluate the $C_{ij}$ values obtained from the optimized parameters for SW and EDIP potentials (see Table \ref{tab:Elast}). For SW Si, we observe that while $C_{12}$ and $C_{44}$ values are closer to DFT than the original potential, $C_{11}$ values are worse. This reveals an inherent limitation of the SW potential wherein optimizing all the $C_{ij}$ parameters together leads to a competing effect. Thus, the optimization performed using the Frobenius norm, which minimizes the error of the sum of all the $C_{ij}$ parameters, results in a Pareto optimal solution. For EDIP Si, similar results are observed where $C_{11}$ and $C_{12}$ values are closer to the DFT solution than the original parameters, while $C_{44}$ is poorer. Interestingly, we observe that the final $C_{ij}$ values obtained by employing the optimized parameters of both EDIP and SW potentials yield similar values. This further confirms the competing nature of $C_{ij}$ values for both the potentials that are being minimized by the Frobenius norm.


\subsection{Vibrational properties}
\label{FC}

\begin{table}
 \caption{Original and optimized parameters for SW and EDIP potential from force constant matching.}
  \centering
  \begin{tabular}{|c|c|c|c|c|} \hline 
    \multirow{2}{*}{Force field params} &\multicolumn{2}{c|}{SW} & \multicolumn{2}{c|}{EDIP}\\ \cline{2-5}
    
     & $\epsilon$ (eV) &  $\lambda$ & A (eV) &  $\lambda$ (eV) \\
    \hline
    Original & 2.16826 & 21.0 & 7.9821730 & 1.4533108 \\
    Optimized - CG& 1.576510867700 & 22.759369692763& 5.710778625018 & 0.746479371508\\
    \hline
  \end{tabular}
  \label{tab:FCparams}
\end{table}

Now, we focus on the vibrational properties computed from the force constants, such as the phonon dispersion and the vibrational density of states. Note that the force constant is computed as the Hessian of the energy. Thus, in this case, we optimize the force-field parameters of SW Si and EDIP Si potentials to minimize the Frobenius norm between the force constants computed using JAX-MD and DFT. Thus, in this case, the inner loop consists of the computation of the force constants and the outer loop consists of backpropagation to minimize the loss function by optimizing the force-field parameters. Thus, in the inner loop Hessian of the energy is computed with respect to the Cartesian coordinates analytically using automatic differentiation to compute the force constants. In the outer loop, to minimize the loss (Eq.~\ref{eq:fro1}) between the computed Hessian and ground truth obtained from DFT, analytical gradients of the Hessian $H_{ij}$ with respect to the model parameters are computed using JAX-MD. This gradients are used to update the force-field parameters iteratively. Since we observed that both BFGS and CG yielded similar results in the case of elastic tensor, here, we employ only the CG optimizer to minimize the loss function. 

Figures~\ref{fig:fig4}(a) and (b) show the loss curves associated with the outer loop iterations for minimizing the error on force constants for SW and EDIP force-field parameters, respectively. Similar to the elastic constant computation, we observe that the loss saturates in 6 and 3 steps for SW and EDIP potentials, respectively. To understand this behavior further, we visualize the loss landscape with respect to the force-field parameters (see Figures~\ref{fig:fig4}(c) and (d)). Similar to the elastic constant, the optimization problem is convex in nature as evidenced by a single global minima in the loss landscape. Further, we project the trajectories of the minimization onto the loss landscape for two different initial conditions. Irrespective of the initial values, the optimization converges rather quickly (in three to four steps) to the global minima. The optimized parameters are shown in Tab.\ref{tab:FCparams}.

\begin{figure}[!htbp]
    \centering
    {\includegraphics[width=0.95\textwidth]{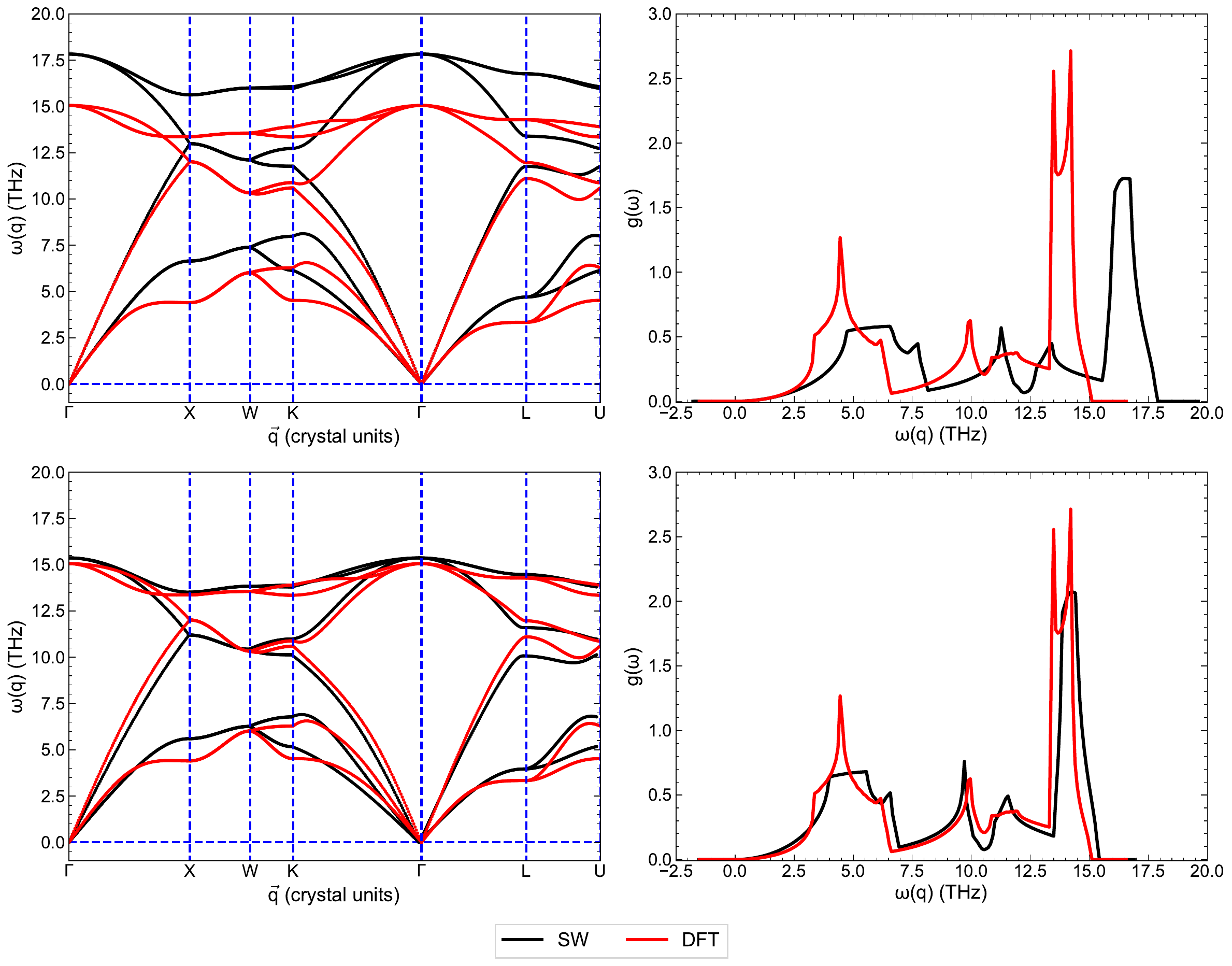}}
    {\includegraphics[width=0.95\textwidth]{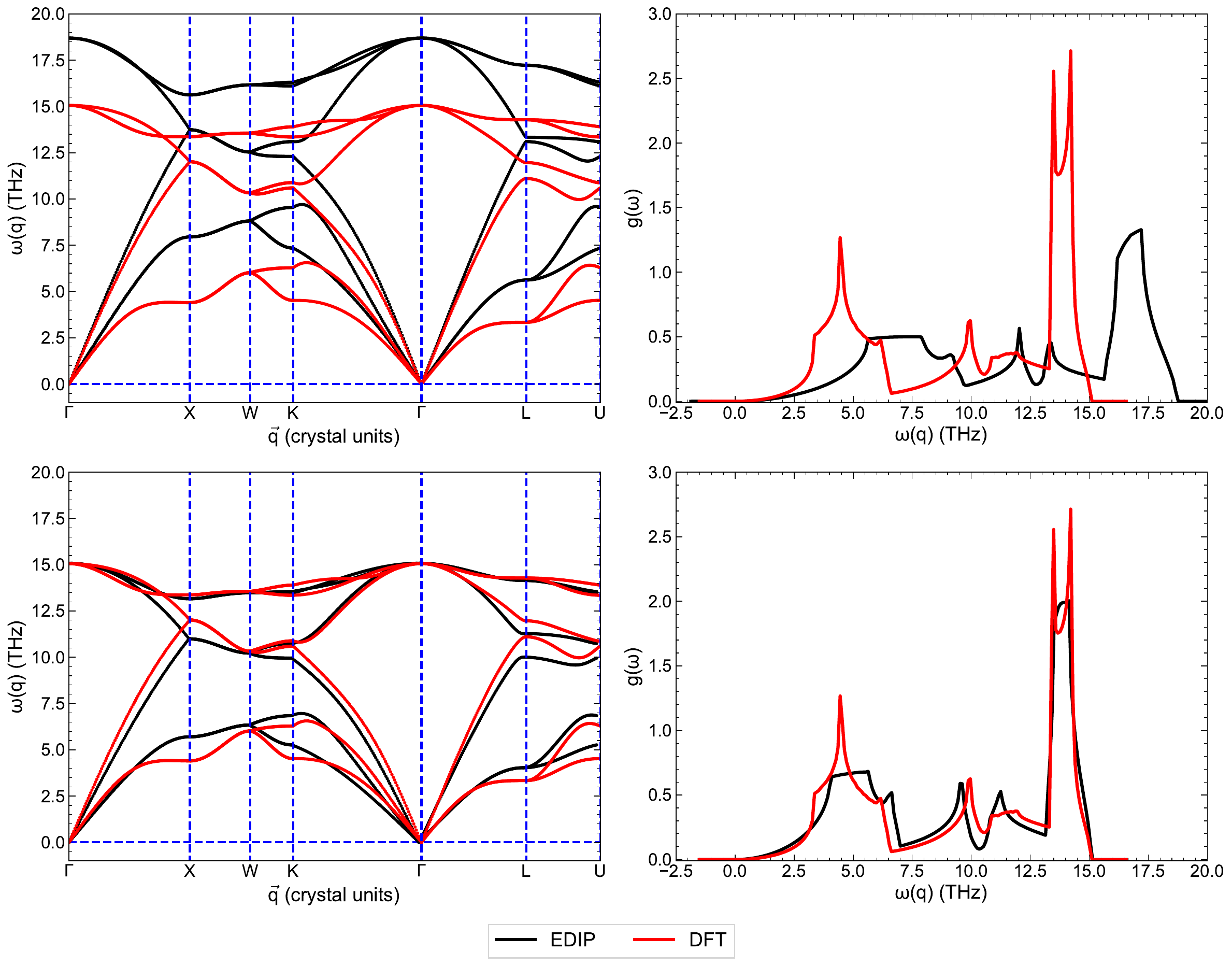}}
\caption{Phonon dispersion and VDOS before (top) and after (bottom) optimization for the SW and EDIP potentials.} 
\label{fig:fig3}
\end{figure}

The optimized force-field parameters are further evaluated by computing the phonon dispersion and vibrational density of states. It is worth noting that these properties can be directly computed from the force constants for which the force-field parameters are optimized. To this extent, the optimized Hessian is passed to the \texttt{phonopy} API in \texttt{python} for post-processing from which the phonon dispersion, vibrational density of states (VDOS), and thermodynamics properties are computed. Figure \ref{fig:fig3} shows the phonon dispersion and vibrational density of states (VDOS) before and after optimization for the SW Si and EDIP Si potential parameters. We observe that for both SW and EDIP potentials, the original parameters exhibit a poor match for VDOS and phonon dispersion. However, the optimized parameters exhibit a much closer match with the VDOS and phonon dispersion obtained from the DFT. It is worth noting that the potential parameters are optimized only for the force constants (Hessian) and not directly on the phonon dispersion or the VDOS. These results further reinforce the capability of automatic differentiation to optimize the force-field parameters toward any target property.

\subsection{Radial distribution function}

\begin{table}
\centering
\caption{Model evaluations for Training and Test data. The energy $E$ and force $F$ errors are RMSE values. The units for energy error are meV / Atom and for force are meV / \AA\ / Atom.}
\label{tab:Models}
    \begin{tabular}{|>{\centering\arraybackslash}p{0.15\linewidth}|c|c|l|l|l|l|c|c|l|l|l|l|} \hline  
         &   \multicolumn{6}{|c|}{Training}&\multicolumn{6}{|c|}{Testing}\\ \hline 
 & \multicolumn{2}{|c|}{RMSE}&  \multicolumn{2}{|c|}{$L^2$}&\multicolumn{2}{|c|}{$L^1$}& \multicolumn{2}{|c|}{RMSE} & \multicolumn{2}{|c|}{$L^2$}& \multicolumn{2}{|c|}{$L^1$}\\
 \hline  
         Models&  $E_{val}$ &  $F_{val}$ &  $E_{val}$& $F_{val}$&$E_{val}$&$F_{val}$& $E_{test}$ &  $F_{test}$  & $E_{test}$ & $F_{test}$  & $E_{test}$ &$F_{test}$  \\ \hline 
         Pre-trained &  4.041& 7.391&  16.33& 54.63&10.26&138.9& 3.877& 7.172& 16.34& 52.66& 10.11&132.8\\ \hline  
         Fine-tuned & 1.034& 3.852&  1.070& 14.84&2.662&74.13& 1.046& 3.745& 1.093& 14.03& 2.635&69.24\\ \hline  
         Fine-tuned (RDF)& 1.070 & 3.834&  1.145& 14.70&2.821&74.05& 1.078& 3.689& 1.162& 13.61& 2.819&68.88\\ \hline 
\end{tabular}
\label{tab:GNNeval}
\vspace{-0.2in}
\end{table}


\begin{figure}[!htbp]
    \centering
    {\includegraphics[width=0.8\textwidth]{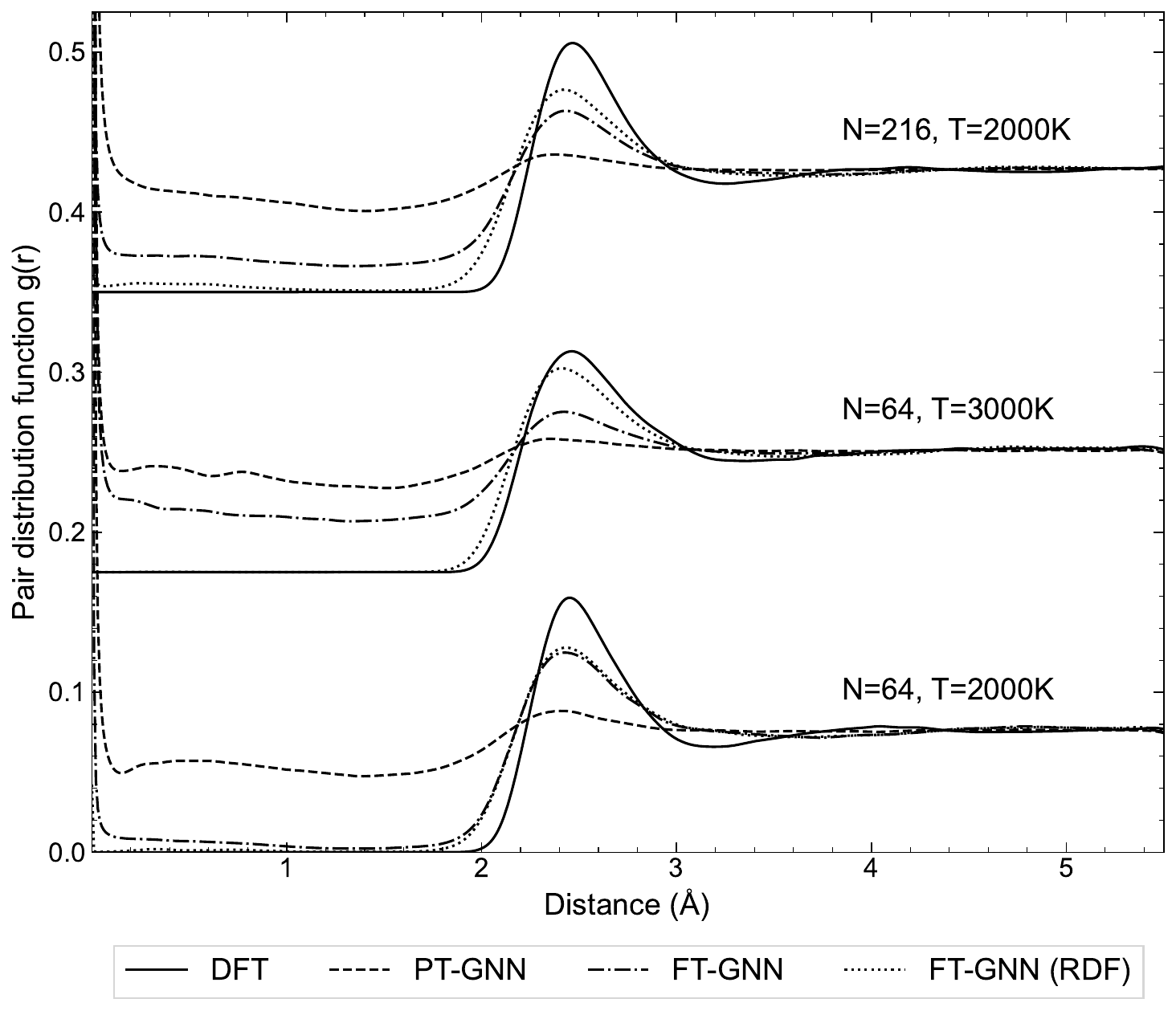}}
\caption{Comparison of the radial distribution function for the pre-trained GNN (PT-GNN), fine-tuned GNN (FT-GNN) and fine-tuned GNN with RDF (FT-GNN (RDF)) with DFT reference. The comparison is done for different particle counts $\mathrm{N}$ and temperature $\mathrm{T}$.} 
\label{fig:fig5}
\end{figure}

Till now, the properties we simulated were obtained based on energy minimization and represented the ground state properties at 0 K. However, most of the properties measurable by experiments are obtained through molecular dynamics simulation. To this extent, we now focus on optimizing the structure of the system at finite temperatures as represented by the radial distribution function (RDF). Note that the RDF captures both the short and medium-range structure of the material and is highly sensitive to the state of the material, that is, solid, liquid, or gas. Without loss of generality, we consider a message-passing graph neural network (MPGNN) pre-trained on the silicon energy and force dataset \cite{SiGNNdataset} as described in \ref{GNNmodel}. We use this model as it is trained on finite temperature data and is expected to perform better than the classical potential. The simulation parameters and the model architecture are given above in section \ref{MDsetup} and \ref{Hyperparameters}. 

The RDF computed from AIMD with 64 silicon atoms at a temperature of 2000 K is used for fine-tuning. The MPGNN is fine-tuned by minimizing the error between the predicted RDF and the ground truth RDF \eqref{eq:RDF}. To test the transferability of this model, we evaluate the RDF at different temperatures, i.e., 1000 K and 3000 K, and a larger system of 216 Silicon atoms at 2000 K. For a more faithful evaluation, we also fine-tune MPGNN model only on the energy and forces at 2000 K. The comparison of the pre-trained (PT-GNN), GNN fine-tuned only on energy and forces (FT-GNN), and GNN fine-tuned on energy and forces along with RDF (FT-GNN (RDF)) is given in figure \ref{fig:fig5} with the quantitative error metrics presented in Tab.~\ref{tab:GNNeval}.

We observe that the PT-GNN is unable to capture the behavior of the Si melt with high non-zero values even for low distances. While FT-GNN gives better performance than PT-GNN, we observe the value of RDF increases significantly closer to 0 distance. This suggests that the model is unable to capture the repulsions at short distances, a well-known problem with MLFFs~\citep{bihani2024egraffbench}. This is further exemplified by the poor performance of FT-GNN at higher temperatures of 3000 K and higher system sizes of 216 atoms. In contrast, we observe that the FT-GNN (RDF) perform significantly better at higher (unseen) temperatures and for larger systems. It is worth noting although the peaks at short distances are absent, the model still exhibits some non-zero values for the RDF at shorter distances. This suggests that the learned repulsive interaction by the FT-GNN (RDF) could be occasionally overcome by high velocity of atoms at high temperatures, a problem that could be addressed by including larger amounts of training data. More importantly, the results demonstrate that automatic differentiation could be used to fine-tune MLFFs or classical potentials towards structural properties obtained from dynamics simulations.

\subsection{Multi-objective optimization}

\begin{figure}[!htbp]
    \centering
    {\includegraphics[width=0.75\textwidth]{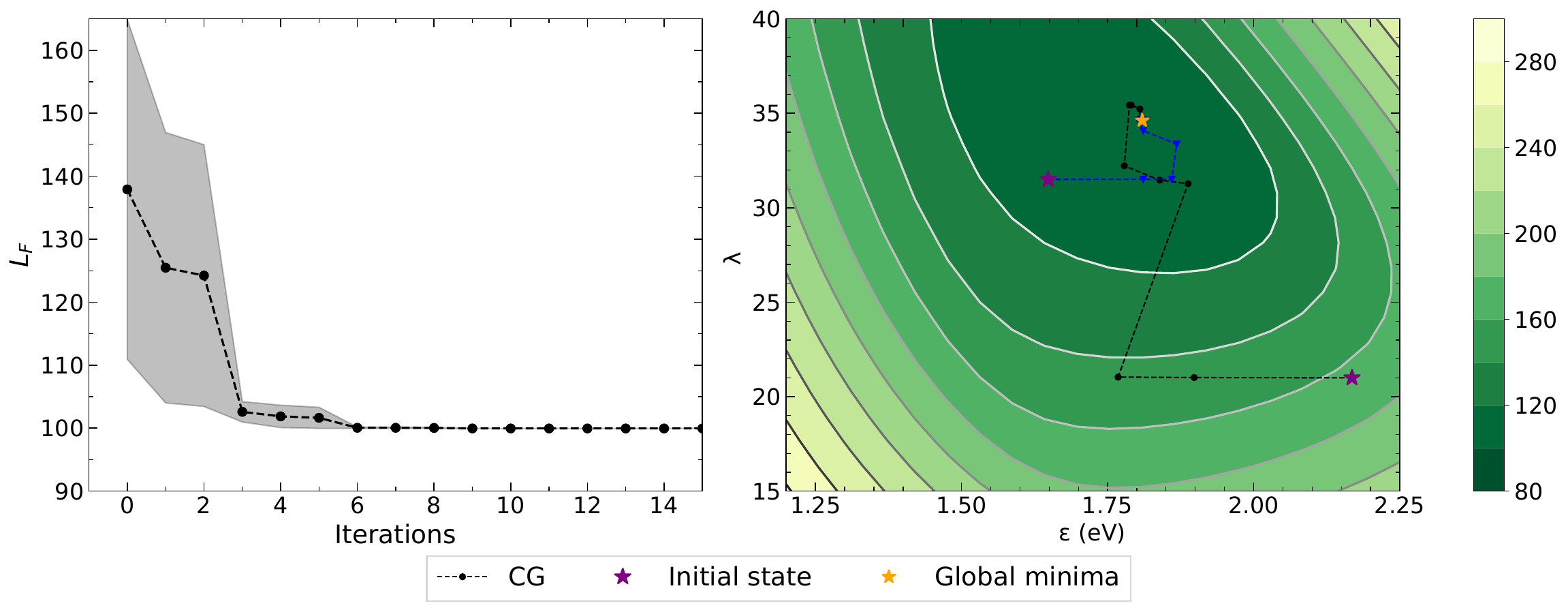}}
    {\includegraphics[width=0.75\textwidth]{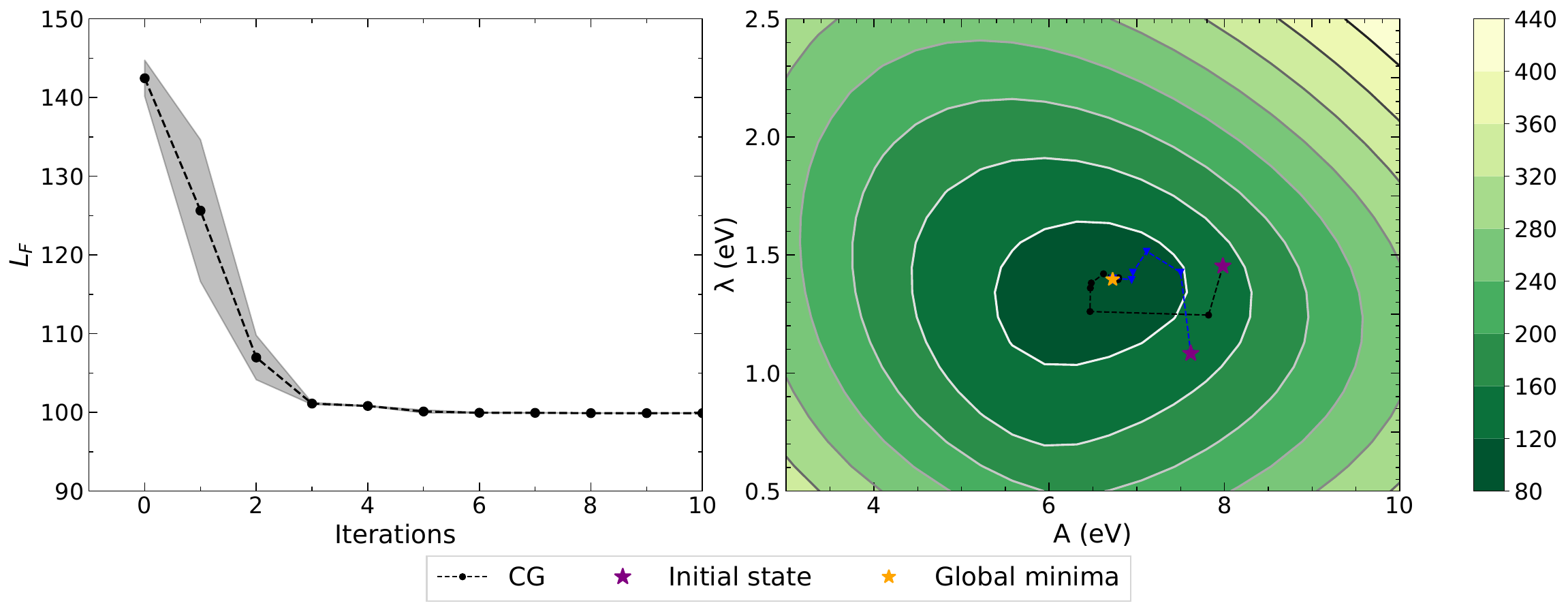}}
\caption{The learning curve and the optimization pathway in the parameter space with CG optimizer for SW (top) and EDIP (bottom) potentials for multi-objective optimization. The contour values are the values of the loss function at each parameter.} 
\label{fig:fig4}
\end{figure}

\begin{table}
 \caption{Original and optimized parameters for SW and EDIP from multi-objective optimization.}
  \centering
  \begin{tabular}{|c|c|c|c|c|} \hline 
    \multirow{2}{*}{Force field params} &\multicolumn{2}{c|}{SW} & \multicolumn{2}{c|}{EDIP}\\ \cline{2-5}
    
     & $\epsilon$ (eV) &  $\lambda$ & A (eV) &  $\lambda$ (eV) \\
    \hline
    Original & 2.16826 & 21.0 & 7.9821730 & 1.4533108 \\
    CG& 1.809324321148&  34.619425564581& 6.723988764513&  1.396547822543\\
    \hline
  \end{tabular}
  \label{tab:Multiparam}
\end{table}

\begin{table}
 \caption{Comparison of elastic constants $C_{ij}$ for original and optimized parameters of SW and EDIP potential along with DFT values for multi-objective optimization.}
  \centering
  \begin{tabular}{|c|c|c|c|c|} \hline 
      \multicolumn{2}{|c|}{Force field} & $C_{11}$ (GPa) &  $C_{12}$ (GPa) & $C_{44}$ (GPa)\\
    \hline
    \multirow{2}{*}{SW} & Original & 151.4131 & 76.4154 &  112.8927 \\
    & Optimized & 153.4054 & 50.2369 & 133.8275 \\
    \hline
    \multirow{2}{*}{EDIP} & Original & 172.0389 & 64.6745 &  145.5683 \\
    &Optimized & 153.4101 & 50.2358 & 133.822 \\
    \hline
    \multicolumn{2}{|c|}{DFT} & 153.2141 & 56.7969 & 149.6653 \\ \hline
  \end{tabular}
  \label{tab:Elastmulti}
\end{table}

Finally, we evaluate the capability of automatic differentiation to optimize force fields on multiple properties. This is particularly useful when several experimental properties of a system are available, all of which could be computed through simulations. Here, we consider optimizing elastic tensor and force constants together. The combined loss function defined in \eqref{eq:fro2}, where $Y_{ij}$ and $\hat{Y}_{ij}$ are matrix elements of $C_{ij}$  while $Z_{ij}$ and $\hat{Z}_{ij}$ are matrix elements of Hessian computed from force fields in JAX-MD and DFT respectively. The weights $w_{1}$ and $w_{2}$ give importance to the property, which is set to 1 in the present case. 

Figure \ref{fig:fig4} shows the loss curve, loss landscape, and the optimization pathway for both SW and EDIP potentials. Similar to the earlier cases, we observe that the global minima are attained in very few iterations (three to six) of the CG minimizer, irrespective of the initialization. Further, we note that the optimized parameters for SW and EDIP potentials are distinctly different from those obtained from single objective optimization (see Tab. \ref{tab:Multiparam} for optimized parameters). This confirms the well-known observation that the parameters optimized from one property are not transferable to reproduce another property. This observation also emphasizes the relevance of multi-objective optimization using differentiable simulation for tuning the potential for target properties. The obtained values of elastic tensors $C_{ij}$ are shown in Tab.~\ref{tab:Elastmulti}. In the case of SW, we observe all the elastic tensors are closer to DFT values for the optimized parameters than the original one. In the case of EDIP, while $C_{11}$ and $C_{12}$ are closer, $C_{44}$ is poorer for the optimized force field. \textcolor{black}{The Phonons and the VDOS resulting from the multi-objective optimization are shown in the supplementary materials.}

\section{Discussion and conclusion}
The extensive experiments presented thus far suggest that AD could be a powerful framework to tune force fields for atomistic simulations even for complex and non-linear properties such as vibrational density of states or radial distribution functions. Indeed, the optimization of pair-wise interactions have been explored in previous studies~\citep{wang2023learning} on LJ and Morse systems by fitting directly on RDFs. The framework presented here extends the approach to any property and systems including MLIPs. Beyond the tuning of force fields, we demonstrate that AD could be used to compute several properties accurately, which can be obtained directly as the gradients of simulations. This represents a paradigm shift from the classical finite difference and other numerical differentiation approaches widely used in atomistic simulations to compute properties. 

The ability of AD to find global minima in a few iterations, albeit for a convex loss landscape, signifies the capability of the AD toward optimizing the force field parameters. Moreover, the simplicity of AD provides a strong test bed for evaluating the flexibility, generalizability, and robustness of force fields, both classical and machine-learned, to simulate complex materials, properties, and phases \citep{Diffthermo}. The fact that the parameters of the force field could be easily optimized to replicate any target property gives a strong handle to build a hypothesis and test what-if scenarios observed experimentally. Further, in the context of universal force fields~\citep{merchant2023scaling,batatia2023foundation}, the approach could be used to either directly fine-tune the potentials to target properties or phases, or to simulate high temperature and pressure configurations or to simply to enhance the stability of the simulations by training on Hessian based properties or RDFs~\citep{raja2024stability}. Moreover, the study demonstrates some of the ``known'' but interesting observations that the classical force fields, due to their limited parameter space, lack the flexibility to replicate multiple properties concurrently, emphasizing the relevance of MLFFs. 

Indeed, the present work could supplement some of the several earlier efforts to build coarse-grained models~\citep{king2024programming,wang2023learning,TorchMD,MultiscaleModellingThesis} and fine-tune potentials on experimental datasets~\citep{thaler2021learning}. AD could be seamlessly used to compute properties in the forward simulation and optimize them against ground truth by computing the automatic gradients. These ground truths could be directly obtained from first-principle simulations, experiments, or even manually constructed by averaging some features. For instance, a major challenge in atomistic simulations is the development of coarse-grained models. To this extent, AD could be used to construct such coarse-grained models by optimizing on a ''synthetic data'' generated to capture the averaged properties from full-atom simulations~\citep{wang2020differentiable,wang2023learning}. Some examples could be the potential for united atom models against the all-atom models or even the bead-string models developed from all-atom models. AD provides a principled way towards attacking these problems ensuring the guarantee of reaching the solution provided it exists.

In conclusion, our study highlights the potential of end-to-end differentiable atomistic simulations for both rapid and precise force field optimization through analytical gradient computation of simulations. The demonstrated applications not only enhance the efficiency of force field refinement but also extend toward computing critical properties, such as derivatives of ensemble averages~\citep{AthermalDesign}. By bridging the gap between machine learning and physics, differentiable simulations provides a fillip to optimization and control in atomistic simulations, setting the stage for broader applications in several areas including insilico materials discovery, self-assembly, topological and material optimization, and enhanced sampling.



\subsubsection*{Acknowledgments}
N.M.A.K acknowledges the funding from Google Research Scholar Award, BRNS YSRA (53/20/01/2021-BRNS), IIT Delhi HPC for the computational and storage resources, M.B. acknowledges the National Science Foundation under Grant No. DMREF- 1922167. \textcolor{black}{This work also used computational and storage services associated with the Hoffman2 Cluster which is operated by the UCLA Office of Advanced Research Computing’s Research Technology Group.}
\bibliography{iclr2024_conference}
\bibliographystyle{iclr2024_conference}

\appendix
\end{document}


\maketitle


\section{Property computation}
\subsection{Elastic constants}

The details of the implementation of the commuting the elastic tensor i.e $C_{ijkl}$ in the framework of athermal elasticity can be found in the \texttt{elasticity.py} module in JAX-MD \citep{AthermalDesign}. 
In Mandel notation, we can write the rank four tensor $C_{ijkl}$ as:

\begin{equation}
    C^{M}_{ij} = \begin{bmatrix}
              C_{1111} & C_{1122} & C_{1133} & \sqrt{2}C_{1123} & \sqrt{2}C_{1113} & \sqrt{2}C_{1112} \\
              C_{2211} & C_{2222} & C_{2233} & \sqrt{2}C_{2223} & \sqrt{2}C_{2213} & \sqrt{2}C_{2212} \\
              C_{3311} & C_{3322} & C_{3333} & \sqrt{2}C_{3323} & \sqrt{2}C_{3313} & \sqrt{2}C_{3312} \\
              \sqrt{2}C_{2311} & \sqrt{2}C_{2322} & \sqrt{2}C_{2333} & 2C_{2323} & 2C_{2313} & 2C_{2312} \\
              \sqrt{2}C_{1311} & \sqrt{2}C_{1322} & \sqrt{2}C_{1333} & 2C_{1323} & 2C_{1313} & 2C_{1312} \\
              \sqrt{2}C_{1211} & \sqrt{2}C_{1222} & \sqrt{2}C_{1233} & 2C_{1223} & 2C_{1213} & 2C_{1212}
              \end{bmatrix}
\end{equation}

All the reported values are in Mandel notation. Note, because of the factors here the $C_{44}$ gets multiplied by a factor of 2 in the Mandel notation. In Voigt notation, the $C_{44}$ is half of the values reported here. Also, because of the symmetry of the diamond cubic crystal, we have only three independent elastic constants. 

In order to check the consistency we compared the values computed using JAX-MD with those from LAMMPS and literature values in table \ref{tab:Elastcomp}
for the SW and EDIP potentials.

\subsection{Force constants}
The simplest way to compute the vibrational properties of a crystal is to use the harmonic approximation whereby we assume that the atoms vibrate around their equilibrium positions. This approximation works when the displacements are small ($\sim 0.01$ \AA) at constant volume. We can perform a Taylor expansion of the potential energy of the crystal as a function of displacements $u_{\alpha}(lp)$: 
\begin{equation}
    U = \Phi_{0} + \sum_{l p} \sum_{\alpha} \Phi_{\alpha}(lp)u_{\alpha}(lp) + \frac{1}{2}\sum_{l p l^{\prime} p^{\prime}} \sum_{\alpha\beta} \Phi_{\alpha\beta}(l p,l^{\prime} p^{\prime})u_{\alpha}(l p)u_{\beta}(l^{\prime} p^{\prime}) + O(u^3)
\end{equation}
Here the labels $l$ and $p$ correspond to the unit cell and the atoms respectively. The ${\alpha,\beta,..}$ correspond to the Cartesian indices, and the $\Phi_{1,2,..,n}$ are the force constants of n\textsuperscript{th} order. We can drop the $O(u^3)$ terms as the displacements are small. At the equilibrium positions $R_{\alpha}(lp)$, the forces $F_{\alpha}(lp)$ are zero, and thus:

\begin{equation}
    \Phi_{\alpha}(lp) = - F_{\alpha}(lp) = \frac{\partial U}{\partial R_{\alpha}(lp)} = 0
\end{equation}

and the second order force constants $\Phi_{\alpha\beta}(l p,l^{\prime} p^{\prime})$ are given as:

\begin{equation}
    \Phi_{\alpha\beta}(l p,l^{\prime} p^{\prime}) = - \frac{\partial  F_{\beta}(l^{\prime}p^{\prime})}{\partial R_{\alpha}(lp)} = \frac{\partial^{2} U}{\partial R_{\alpha}(lp) \partial R_{\beta}(l^{\prime} p^{\prime})}
\end{equation}

Since we do not know the analytic force constants they are calculated by applying a finite displacement (FD) to atoms and calculating the resulting forces on other atoms as:

\begin{equation}
\label{eq:FD}
    \Phi_{\alpha\beta}(l p,l^{\prime} p^{\prime}) \simeq - \frac{F_{\beta}(l^{\prime}p^{\prime},\Delta R_{\alpha}(lp)) - F_{\beta}(l^{\prime}p^{\prime})}{\Delta R_{\alpha}(lp)}
\end{equation}

In theory, the force constant matrix (hessian) mentioned above should be computed for an infinite lattice however in practice we use another approximation that the induced forces due to displacements of atoms far away are small or close to zero (Frozen phonon approximation). For $N$ atoms this leaves us with a total of $3N$ force calculations or $6N$ if the centering difference method is used. For materials where the force constants decay slowly as a function of the super-cell size, it becomes computationally challenging to perform a large number of force calculations, especially when using \textit{ab initio} methods like DFT. Hence, modern packages like \texttt{phonopy} \citep{phonopy} use symmetry to reduce the number of force calculations and also improve numerical stability. 

Once the force constant matrix is calculated we can set up the dynamical matrix in the reciprocal space:

\begin{equation}
    D^{\alpha \beta}_{p p^{\prime}}(q) = \sum_{l^{\prime}}\frac{\Phi_{\alpha \beta}(0p,l^{\prime}p^{\prime})}{\sqrt{m_{p}m_{p^{\prime}}}}\exp({iq\cdot[R(l^{\prime}p^{\prime}) - R(0p)])}
\end{equation}

where $q$ is the phonon wave vector and $m_{p}$ is the mass of the atom $p$. With the approximations made, we can set up an eigenvalue problem to calculate the phonon frequencies $\omega_{q j}$ and the phonon eigenvectors $e_{q j}$ as follows:

\begin{equation}
    \sum_{\beta p^{\prime}}D^{\alpha \beta}_{p p^{\prime}}(q)e^{\beta p^{\prime}}_{q j} = \omega^{2}_{q j}e^{\alpha p}_{q j}
\end{equation}

Here $j$ is the band index. For a 3-dimensional system, the dynamical matrix is $3N\times 3N$ hermitian. The phonon density of states (PDOS) then can be calculated as:

\begin{equation}
    g(\omega) = \frac{1}{l_{max}}\sum_{q j} \delta(\omega - \omega_{q j})
\end{equation}

The $l_{max}$ is the number of unit cells used. This ensures that $\int g(\omega)d\omega = 3N$.
Using the obtained phonon frequencies we can compute the phonon number $n$ and the energy of the phonon system $E_{ph}$ under the harmonic approximation:

\begin{equation}
    n = \frac{1}{\exp(\frac{\hbar\omega_{qj}}{k_{B}T}) - 1}
\end{equation}
\begin{equation}
    E_{ph} = \sum_{qj}\hbar\omega_{qj}\left[\frac{1}{2} + \frac{1}{\exp(\frac{\hbar\omega_{qj}}{k_{B}T}) - 1}  \right]
\end{equation}

Where $\hbar, k_{B}$ and $T$ are the Plank's constant, Boltzmann constant, and the temperature respectively. Using the known relations in statistical mechanics we can further compute the heat capacity at constant volume $(C_{v})$:

\begin{equation}
\label{cv}
    C_{v} = \left (\frac{\partial E}{\partial T} \right)_{V} = \sum_{qj} k_{B} \left (\frac{\hbar\omega_{qj}}{k_{B}T} \right)^2\frac{\exp(\frac{\hbar\omega_{qj}}{k_{B}T})}{[\exp(\frac{\hbar\omega_{qj}}{k_{B}T}) - 1]^2}
\end{equation}

Using the canonical partition function $Z$ the free energy $F$ and entropy $S$ can be computed:

\begin{equation}
    Z = \exp(\frac{-\psi}{k_{B}T})\prod_{qj} \frac{\exp(\frac{-\hbar\omega_{qj}}{2k_{B}T})}{1 - \exp(\frac{-\hbar\omega_{qj}}{k_{B}T})}
\end{equation}

\begin{equation}
\label{f}
    F = -k_{B} T \ln{Z} = \psi + \frac{1}{2}\sum_{qj}\hbar \omega_{qj} + k_{B}T\sum_{qj}\ln[1 - \exp(\frac{-\hbar\omega_{qj}}{k_{B}T})]
\end{equation}

\begin{equation}
\label{s}
    S = -\frac{\partial F}{\partial T} = \frac{1}{2T}\sum_{qj}\hbar\omega_{qj}\coth({\frac{\hbar\omega_{qj}}{2k_{B}T}}) - k_{B}\sum_{qj}\ln[2\sinh({\frac{\hbar\omega_{qj}}{2k_{B}T}})]
\end{equation}

In JAX-MD we have access to analytical gradients using AD and thus we do not have to perform any approximation in computing the force constants like in eq \ref{eq:FD}. This can be of particular advantage as the choice of displacement can alter the results. A large displacement will mean that anharmonicities are introduced while a small displacement will underestimate the curvature. AD force constants are also more numerically stable. Post-processing of both force constants is done by \texttt{phonopy} \citep{phonopy} as mentioned before.

\begin{figure}[!htbp]
    \centering
    {\includegraphics[width=0.95\textwidth]{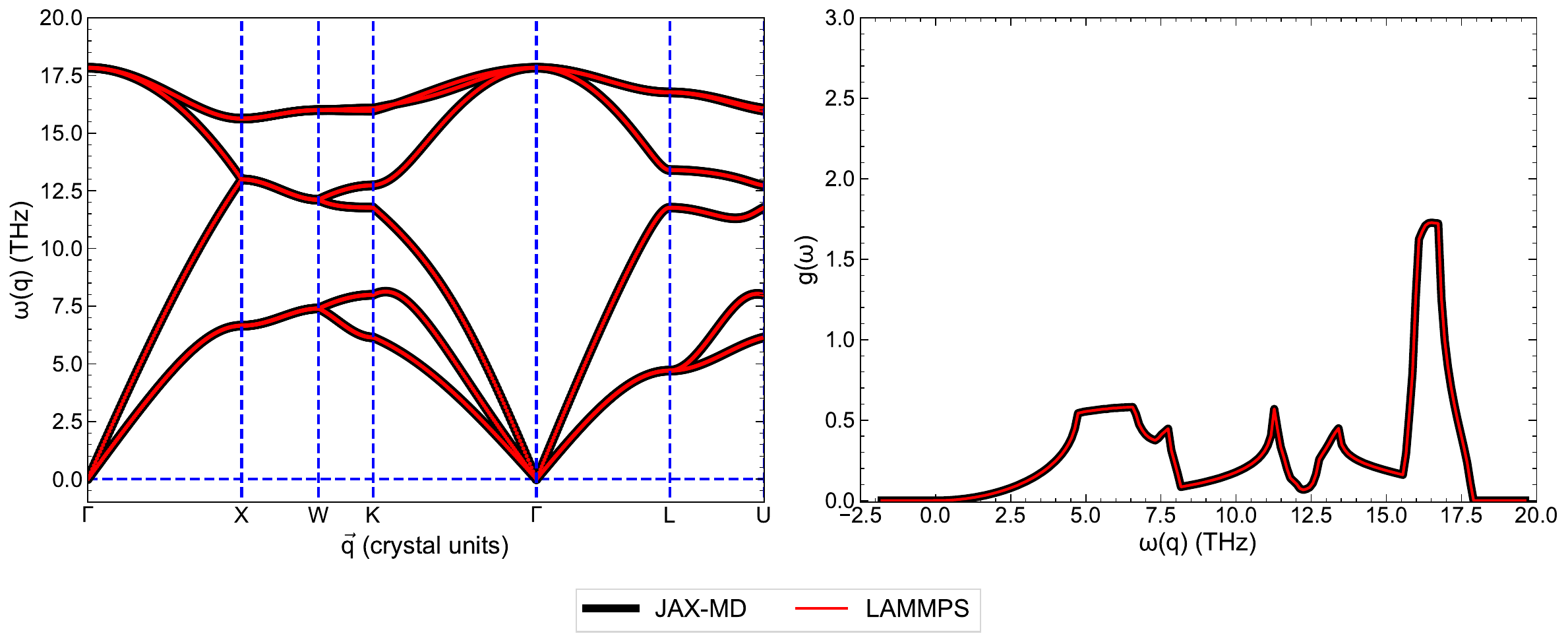}}
    {\includegraphics[width=0.95\textwidth]{Figures_supplementary/Phonon_comp_LAMMPS_SW.pdf}}
\caption{Phonon dispersion and VDOS calculated from LAMMPS and JAX-MD for the SW (top) and EDIP (bottom) potentials.} 
\label{fig:LAMMPS_comp}
\end{figure}

We first compared the two methods for Silicon in the diamond phase by comparing the phonon dispersion and the PDOS as shown in figure \ref{fig:LAMMPS_comp}. The force constants produced by the two methods yield identical results. Note that the AD force constants calculated in JAX-MD do not consider any symmetry while those calculated with LAMMPS do.

\subsection{RDF}
One further application of differentiable molecular dynamics would be to optimize the force-field parameters to match the distributions obtained from the AIMD simulation. We consider the radial distribution function (RDF) computed from AIMD simulation. The RDF is defined as:

\begin{equation}
    g(r) = \left<\sum_{i \neq j}\delta(r - |r_i - r_j|)\right>
\end{equation}

We make an approximation that:

\begin{equation}
     \delta(r) \approx \exp\left( -\frac{1}{2} \frac{(dr - r)^2}{\sigma^2} \right) \left( \frac{1}{r + \epsilon} \right)^{d - 1}
\end{equation}
where $dr$ is the distance between particles and $d$ is the dimension of the space. $\epsilon$ is a small parameter for stability for $r=0$. We set $\sigma = 0.05$\ \AA\ and the radius cutoff for the RDF to be 5.5 \AA. The implementation can be found in the \texttt{quantity.pair\_correlation} and \texttt{quantity.pair\_correlation\_neighbor\_list} functions in JAX-MD.

\textbf{Code snippet:} Code for optimizing the force field for target elastic properties.
\begin{lstlisting}[language=Python]
import jax.numpy as jnp
from jax import grad, jit
from jax_md import space, energy, quantity, elasticity
    
# Coordinates
R = jnp.array([[0.25, 0.75, 0.25],
       [0.  , 0.  , 0.5 ],
       [0.25, 0.25, 0.75],
       [0.  , 0.5 , 0.  ],
       [0.75, 0.75, 0.75],
       [0.5 , 0.  , 0.  ],
       [0.75, 0.25, 0.25],
       [0.5 , 0.5 , 0.5 ]])

# Lattice vector
latvec = jnp.array([[5.43000, 0.00000, 0.00000], 
                    [0.00000, 5.43000, 0.00000],
                    [0.00000, 0.00000, 5.43000]])

# Parameters of forcefield
params = jnp.array([2.16826, 21.0])

# Reference elastic tensor
C_t = jnp.array([[153.21414,56.79690,56.79690,0,0,0],
                 [56.79690,153.21414,56.79690,0,0,0],
                 [56.79690,56.79690,153.21414,0,0,0],
                 [0,0,0,149.66528,0,0],
                 [0,0,0,0,149.66528,0],
                 [0,0,0,0,0,149.66528]])

def setup(R, latvec, C_t):
  # Setup simulation environment
  displacement, shift = space.periodic_general(latvec)
  dist_fun = space.metric(displacement)

  @jit
  def run(params):
    # Function to calculate the elastic tensor

    # Define the energy function
    energy_fn = energy.stillinger_weber(displacement,epsilon=params[0], lam=params[1])

    # Solve for elastic tensor
    emt_fn = elasticity.athermal_moduli(energy_fn)
    C = emt_fn(R, latvec)
    C_p = jit(elasticity.tensor_to_mandel)(C)

    # Return the error between predicted and true elastic tensor 
    # based on Frobenius Norm
    return jnp.linalg.norm(C_p - C_t, ord='fro')
  return run

run = setup(R, latvec, C_t)

# Computes the elastic tensor for given params
run(params)

# Computes the gradient of elastic tensor w.r.t to params
grad(run)(params)
\end{lstlisting}

\newpage
\begin{table}
 \caption{Comparison of elastic constants $C_{ij}$ obtained from JAX-MD with LAMMPS and Reference values with the SW \citep{SW_analytical} and EDIP potential \citep{EDIP2}.}
  \centering
  \begin{tabular}{|c|c|c|c|c|} \hline 
      \multicolumn{2}{|c|}{Force field} & $C_{11}$ (GPa) &  $C_{12}$ (GPa) & $C_{44}$ (GPa)\\
    \hline
    \multirow{2}{*}{SW} & JAX-MD& 151.4131 & 76.4154 &  112.8927 \\
    & LAMMPS& 151.4244 & 76.4220 & 112.89866 \\
 & Literature  & 151.42 & 76.42 & 112.9 \\
    \hline
    \multirow{2}{*}{EDIP} & JAX-MD& 172.0389 & 64.6745 &  72.7842 \\
    &LAMMPS& 171.9912 & 64.7164 & 72.7521  \\
 & Literature& 175 &  62 &  71\\
 \hline
  \end{tabular}
  \label{tab:Elastcomp}
\end{table}

\newpage
\begin{figure}[!htbp]
    \centering
    {\includegraphics[width=0.475\textwidth]{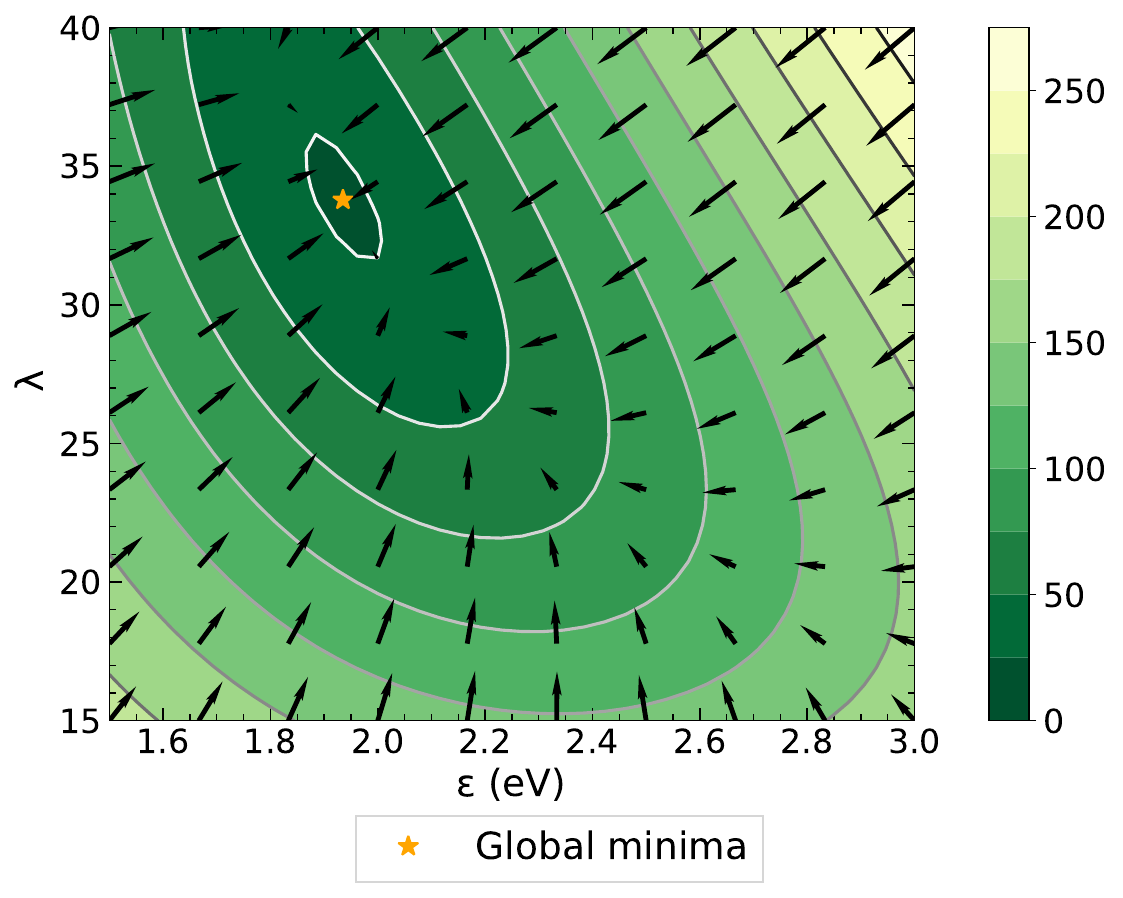}}
    {\includegraphics[width=0.475\textwidth]{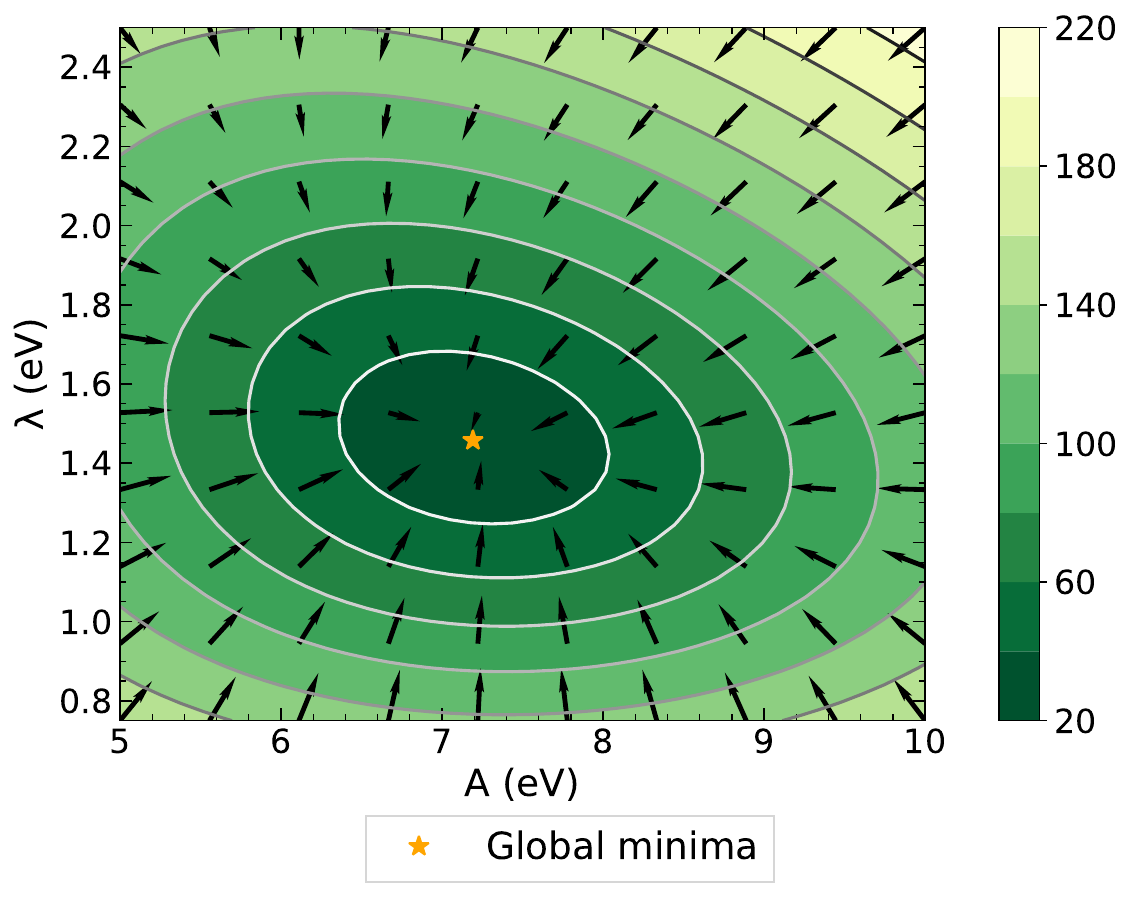}}
\caption{Contour plot of the loss function in the case of elastic constant optimization for SW (left) and EDIP (right). The black arrows represent the gradient direction and magnitude.} 
\label{fig:contour_grad}
\end{figure}

\newpage
\begin{figure}[!htbp]
    \centering
    {\includegraphics[width=0.475\textwidth]{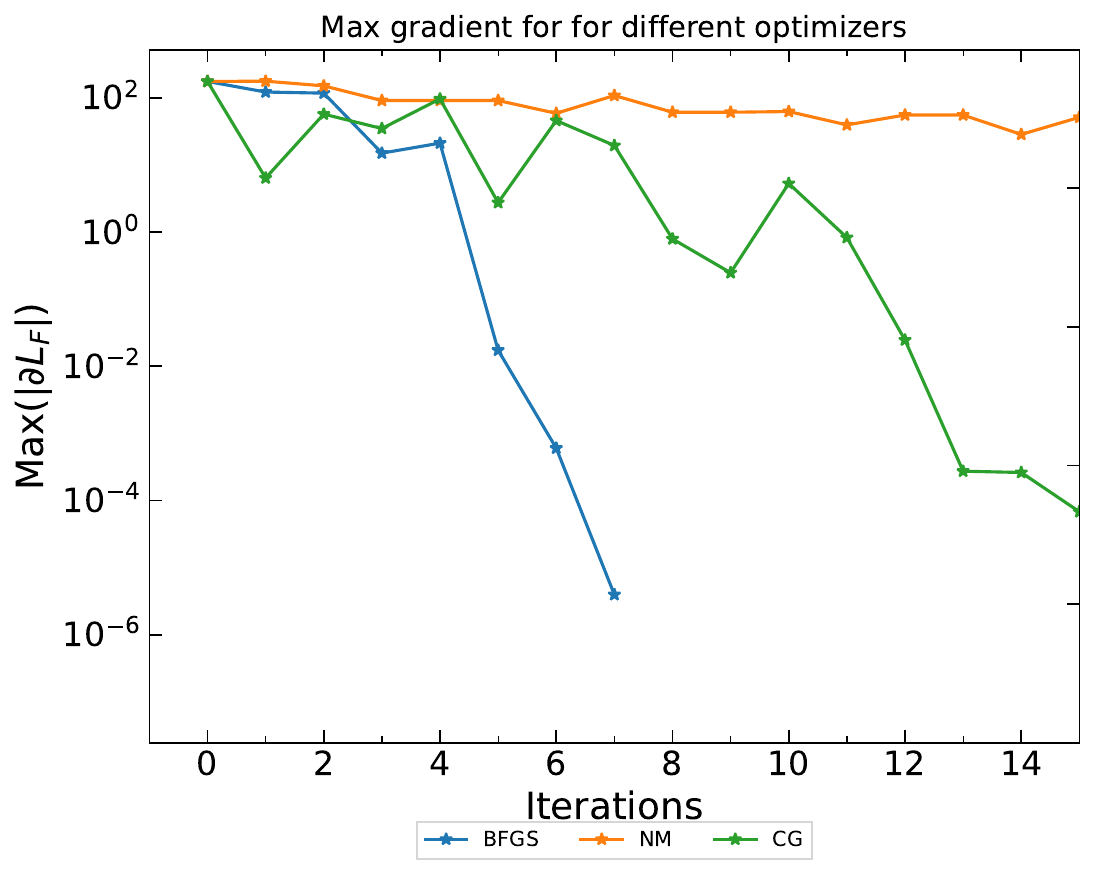}}
    {\includegraphics[width=0.475\textwidth]{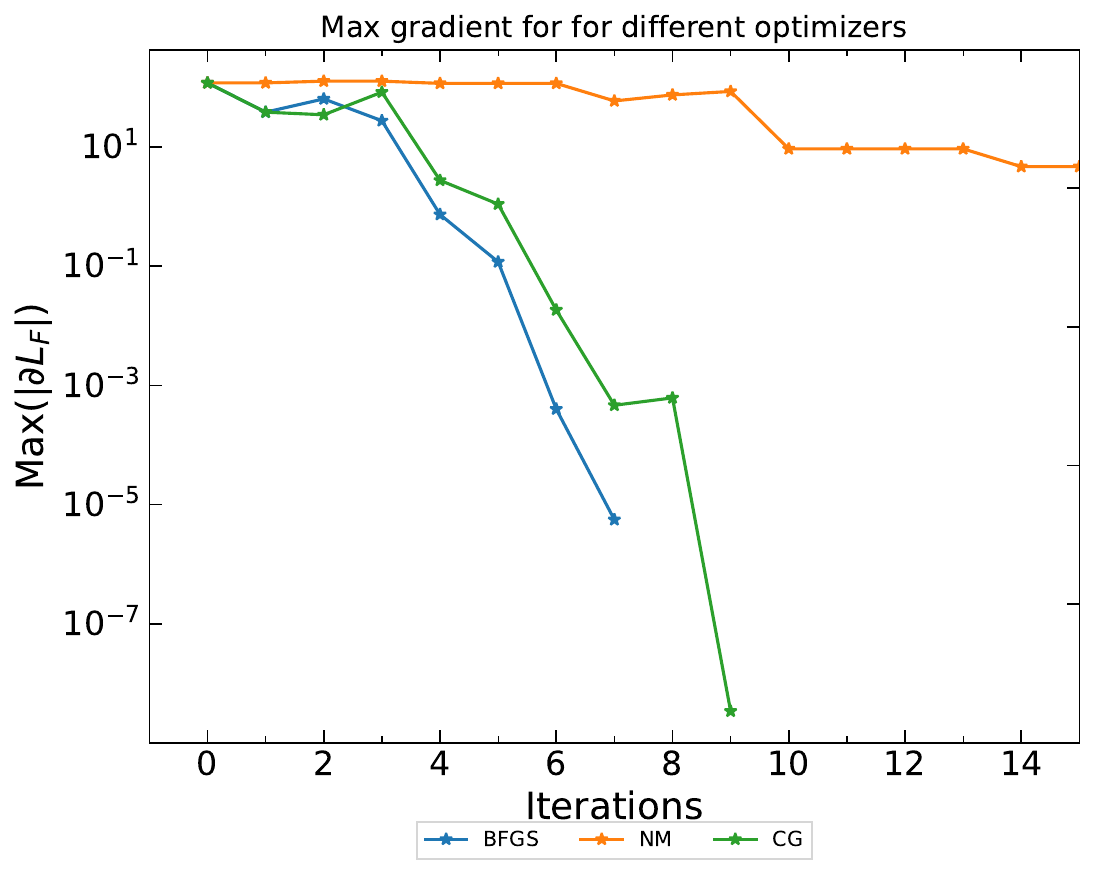}}
\caption{Comparison of maximum value of the gradient at each step of different optimizers for elastic constant optimization for the SW (left) and EDIP (right) potentials.} 
\label{fig:max_grad}
\end{figure}

\newpage

\newpage
\begin{figure}[!htbp]
    \centering
    {\includegraphics[width=0.95\textwidth]{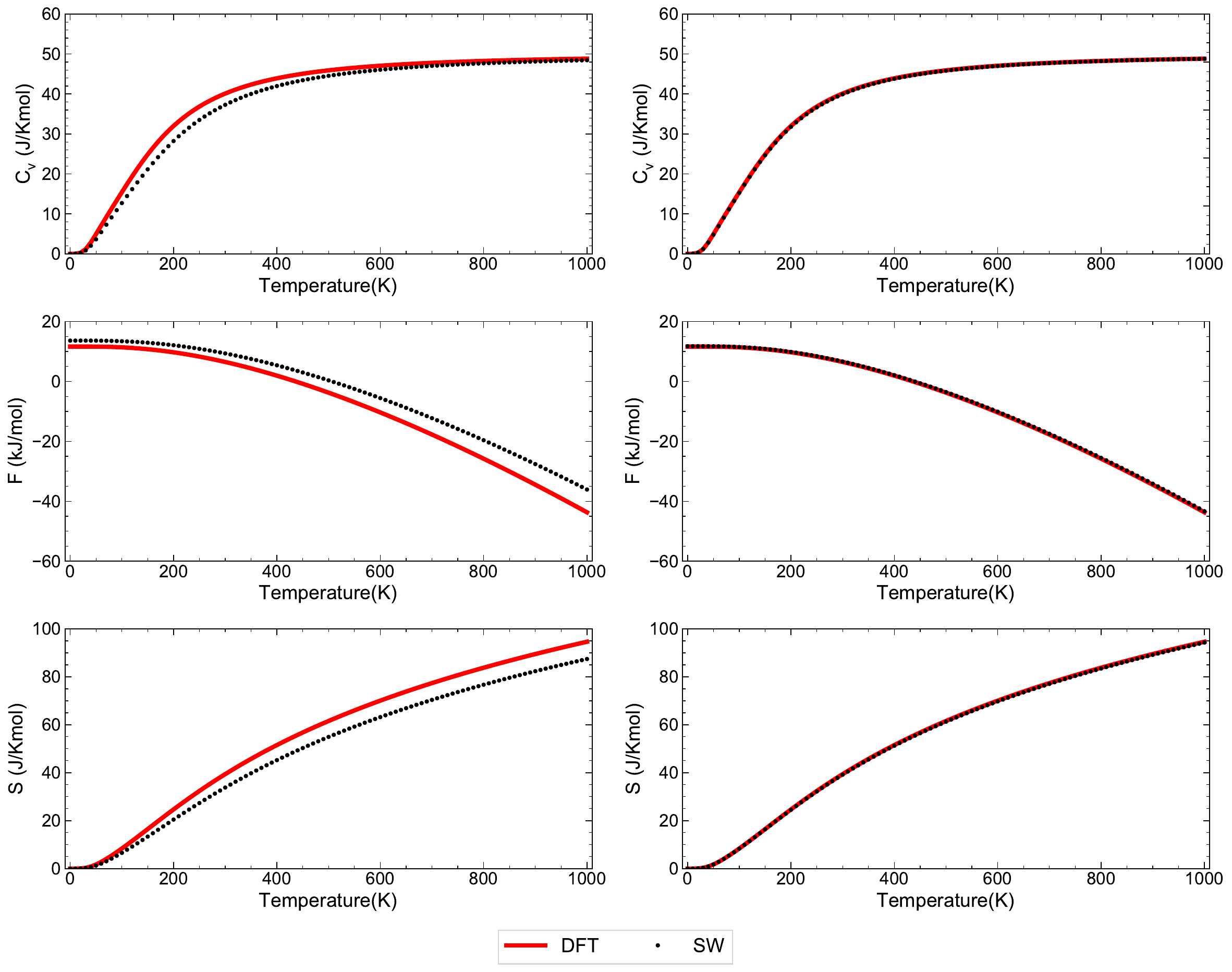}}
\caption{Thermodynamic properties as computed from \cref{cv,f,s} before (left) and after (right) force constant matching for the SW potential.} 
\label{fig:thermo_SW}
\end{figure}

\newpage
\begin{figure}[!htbp]
    \centering
    {\includegraphics[width=0.95\textwidth]{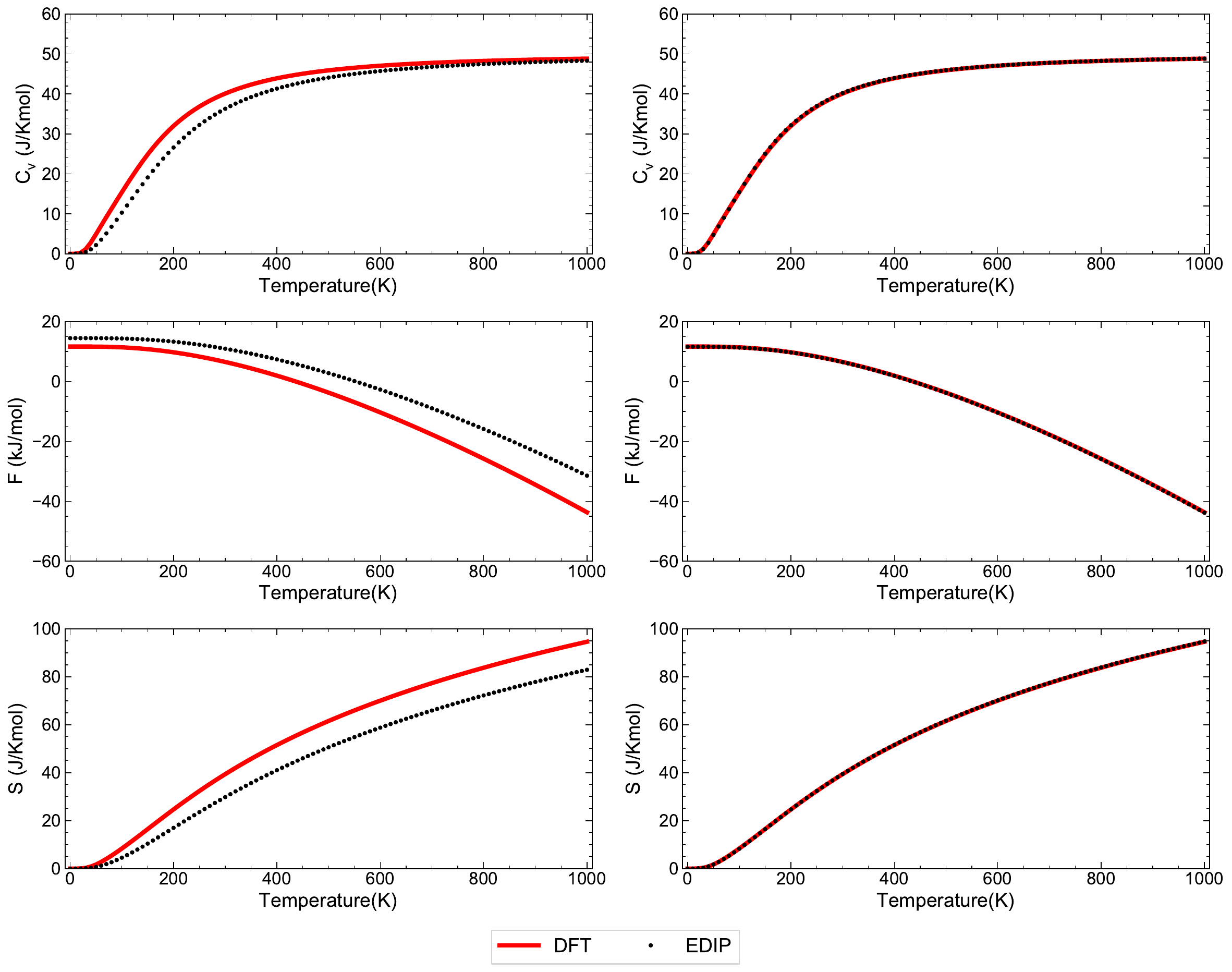}}
\caption{Thermodynamic properties as computed from \cref{cv,f,s} before (left) and after (right) force constant matching for the EDIP potential.} 
\label{fig:thermo_EDIP}
\end{figure}

\newpage
\begin{figure}[!htbp]
    \centering
    {\includegraphics[width=0.95\textwidth]{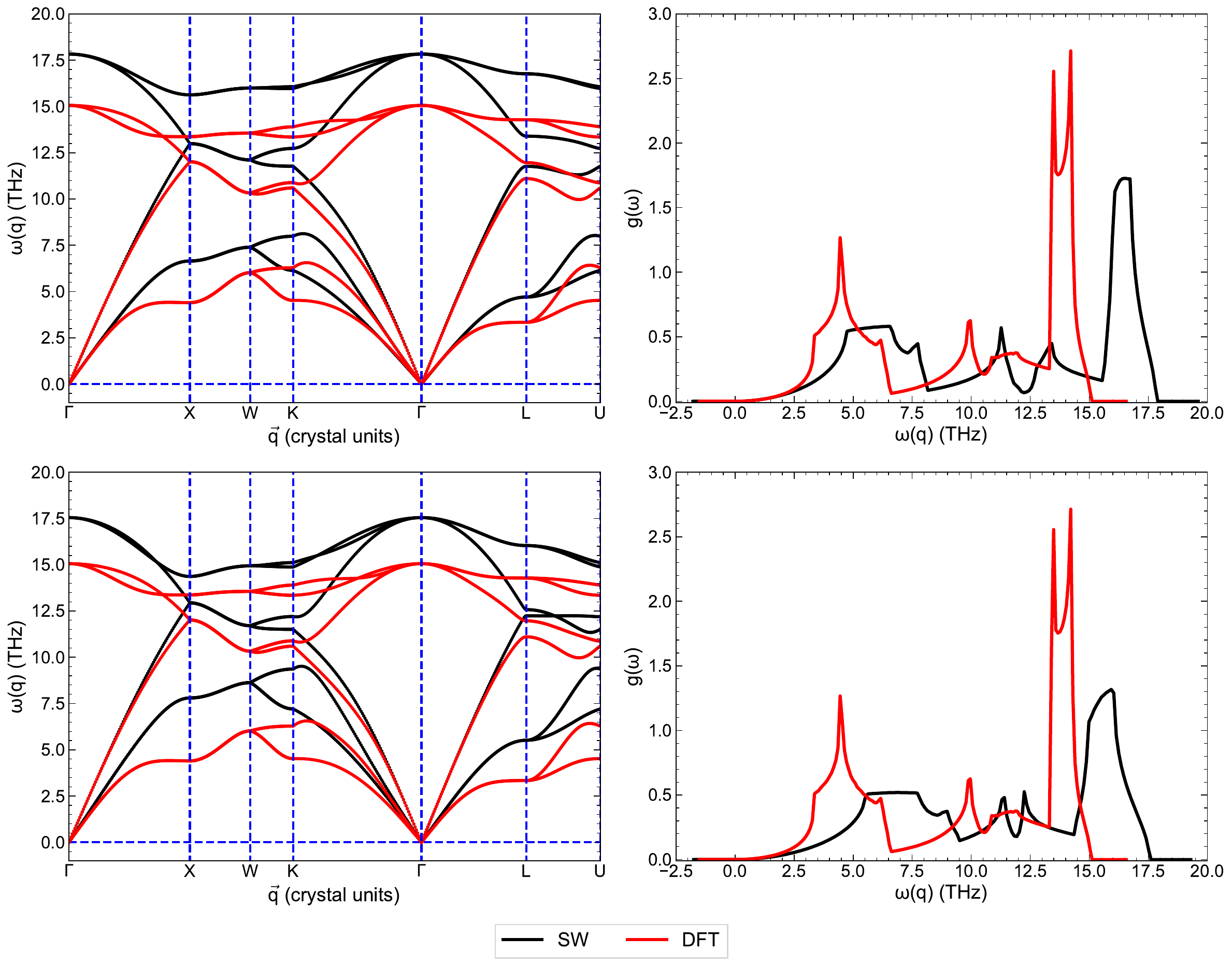}}
    {\includegraphics[width=0.95\textwidth]{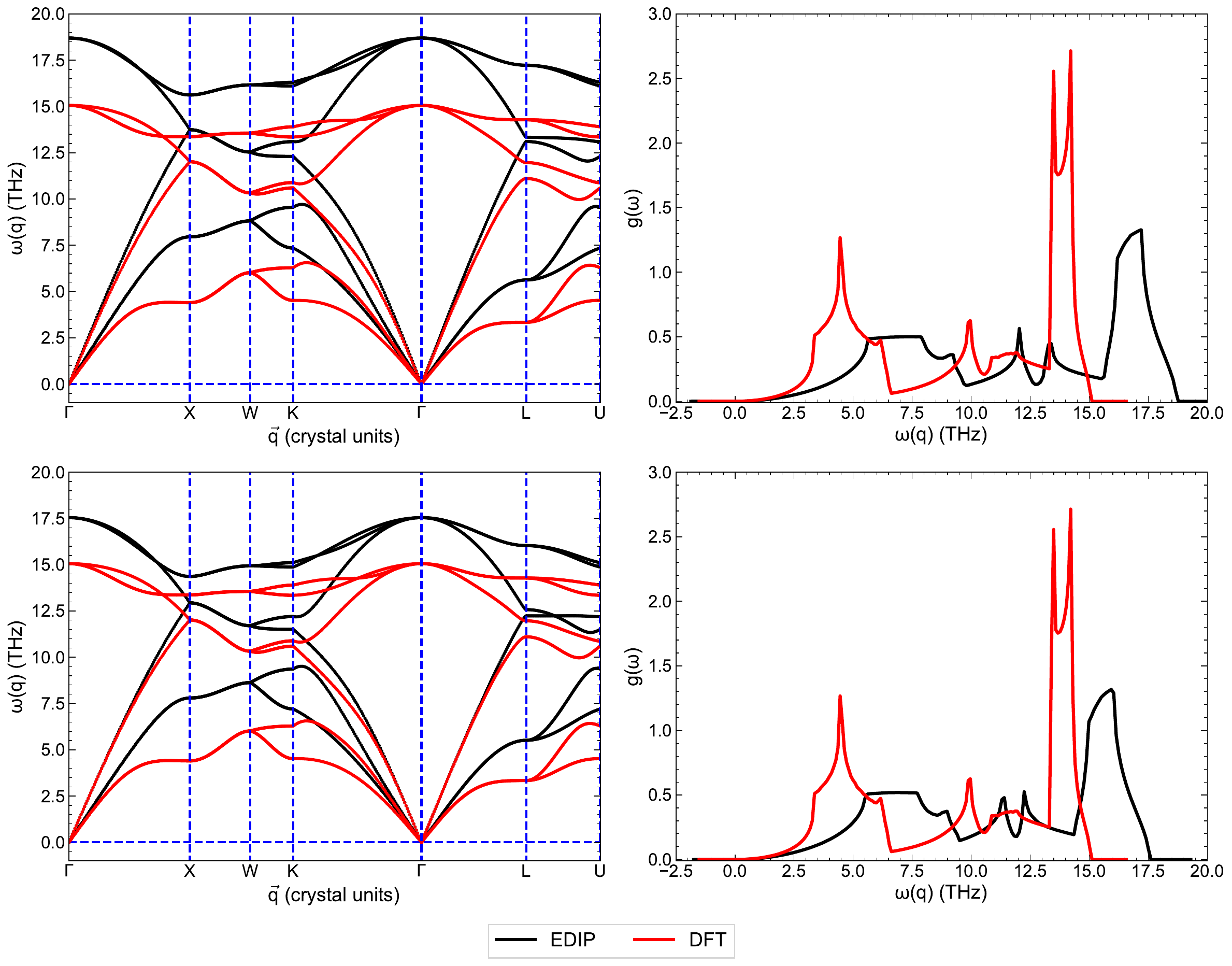}}
\caption{Phonon dispersion and VDOS before (top) and after (bottom) optimization for the SW and EDIP potentials for multi-objective optimization.} 
\label{fig:Multi_comp}
\end{figure}

\newpage
\begin{figure}[!htbp]
    \centering
    {\includegraphics[width=0.95\textwidth]{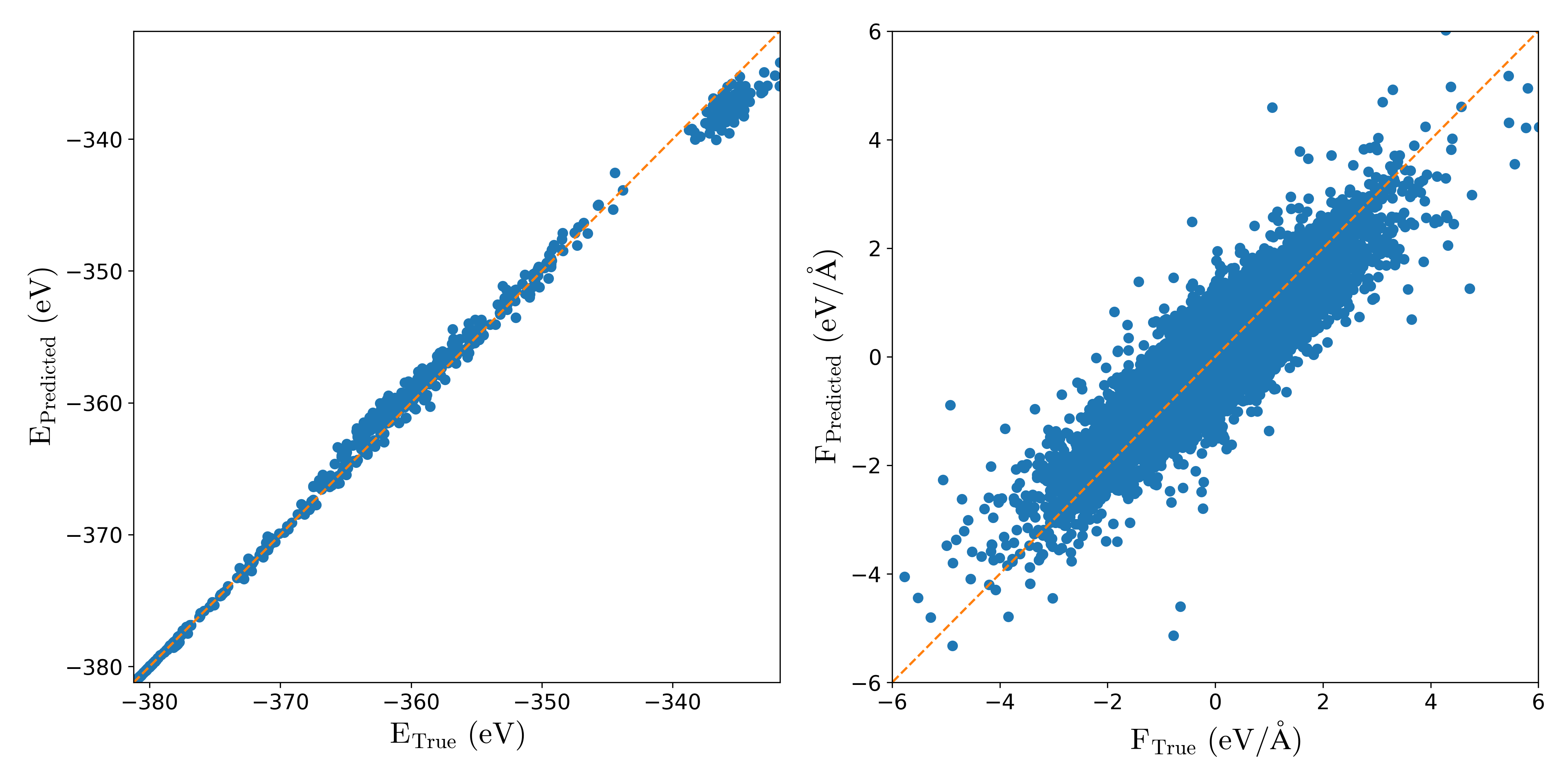}}
    {\includegraphics[width=0.95\textwidth]{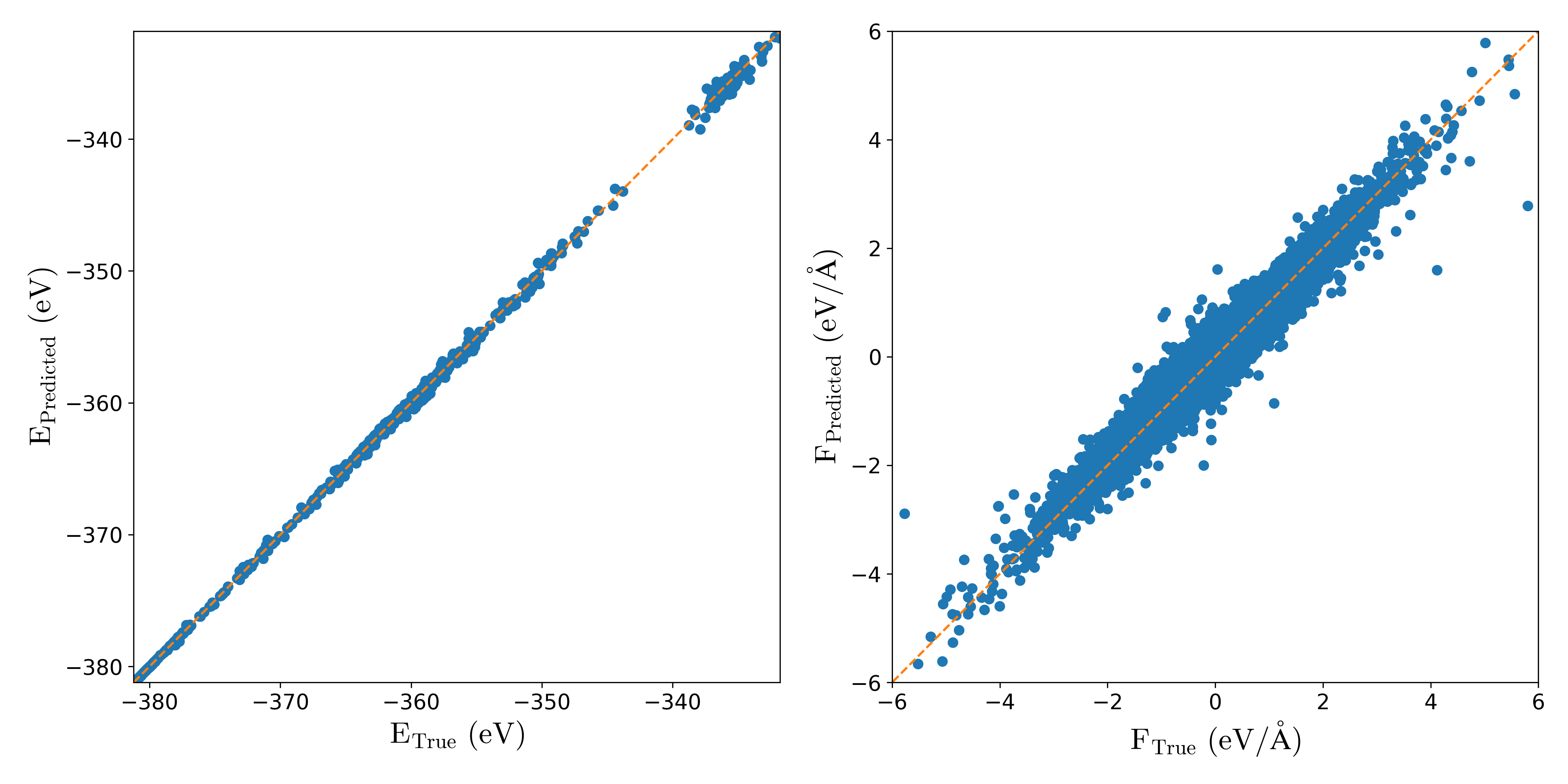}}
    {\includegraphics[width=0.95\textwidth]{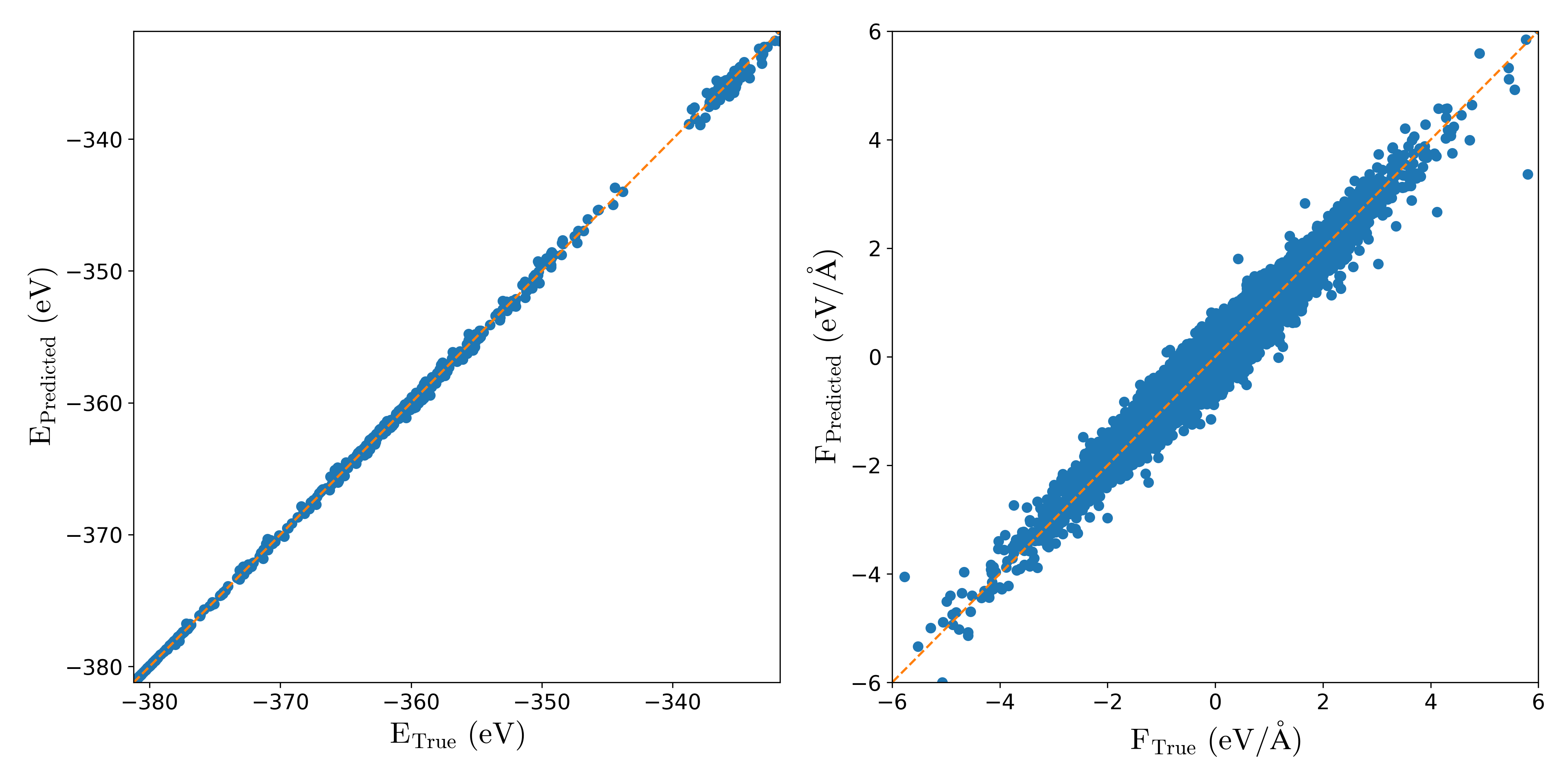}}
\caption{Comparison of the true vs predicted energy and force for the PT-GNN (top), FT-GNN (middle), and FT-GNN (RDF) bottom.} 
\label{fig:pred_true_plot}
\end{figure}

\newpage
\bibliography{iclr2024_conference}
\bibliographystyle{iclr2024_conference}